\documentclass[a4paper,12pt]{JHEP3}
\usepackage{epsfig,multicol,cite}

\newcommand{\newc}{\newcommand}
\newc{\tev}{\,{\rm TeV}}
\newc{\gev}{\,{\rm GeV}}
\newc{\sgn}{\mr{sgn}\,}
\newc{\ra}{\rightarrow}
\newc{\rpv}{$\mathrm{\not\!R_p}$}
\newc{\met}{$\not\!\!E_T$}
\newc{\rp}{$\mathrm{R_p}$}
\newc{\real}{\mathcal{R}e}
\newc{\alsm}{{\displaystyle \sum_{\alpha=1,2}}}
\newc{\besm}{{\displaystyle \sum_{\beta=1,2}}}
\newc{\al}{\alpha}
\newc{\ga}{\gamma}
\newc{\de}{\delta}
\newc{\cw}{\cos\theta_w}
\newc{\ssw}{\sin^2\theta_w}
\newc{\ccw}{\cos^2\theta_w}
\newc{\cbe}{\cos\beta}
\newc{\sbe}{\sin\beta}
\newc{\sh}{\hat{s}}
\newc{\sa}{\sin\al}
\newc{\ca}{\cos\al}
\newc{\bv}{$\mathrm{\not\!B}$}
\newc{\lv}{$\mathrm{\not\!L}$}
\newc{\ie}{{\it i.e.\/}\ }
\newc{\lam}{\lambda}
\newc{\cht}{\tilde{\chi}}
\newc{\upt}{\tilde{u}}
\newc{\elt}{\tilde{\ell}}
\newc{\hgt}{\tilde{H}}
\newc{\nut}{\tilde{\nu}}
\newc{\dnt}{\tilde{d}}
\newc{\psb}{\bar{\psi}}
\newc{\rtt}{\sqrt{2}}
\newc{\mut}{\tilde{\mu}}
\newc{\mr}{\mathrm}
\newc{\bath}{\bar{\theta}}
\newc{\tht}{\theta}
\newc{\JC}{{\bf J}}
\newc{\lra}{\longrightarrow}
\newc{\eg}{{\it e.g.\,}}
\newc{\barr}{\begin{array}}
\newc{\earr}{\end{array}}
\newc{\dis}{\displaystyle}
\newc{\beq}{\begin{equation}}
\newc{\eeq}{\end{equation}}
\newc{\me}{\mathcal{M}}
\newc{\dbm}{\partial_\mu}
\newc{\sgm}{\sigma_\mu}

\def\ra{\rightarrow}
\def\dis{\displaystyle}
\catcode`@=11 
\def \gsim{\mathrel{\mathpalette\@versim>}}
\def \lsim{\mathrel{\mathpalette\@versim<}}
\def \@versim#1#2{\lower0.4ex\vbox{\baselineskip\z@skip\lineskip\z@skip
     \lineskiplimit\z@\ialign{$\m@th#1\hfil##\hfil$%
     \crcr#2\crcr\sim\crcr}}}
\catcode`@=12 
\def\gev{\: \rm GeV}

\renewcommand{\thefootnote}{\fnsymbol{footnote}}

\title{ NLO Corrections to lepton pair production beyond the 
Standard Model at hadron colliders}
\author{Debajyoti Choudhury\footnote{E-mails: debchou@physics.du.ac.in} \\
 Department of Physics and Astrophysics, University of Delhi,\\ 
       Delhi 110 007, India.}

\author{Swapan Majhi\footnote{E-mails:  swapan\_majhi@baylor.edu}\\
Department of Physics, Baylor University, Waco, TX 76706, U.S.A.}

 \author{V. Ravindran\footnote{E-mails: ravindra@mri.ernet.in}\\
Harish-Chandra Research Institute, Chhatnag Road, Jhusi,\\ 
      Allahabad 211 019, India. }
\abstract{
We consider lepton pair production at a hadron collider in a class 
of effective theories with the relevant operators being four-fermion 
contact interaction. Despite the nonrenormalizable nature of the interaction, 
we explicitly demonstrate that calculating QCD corrections is both possible 
and meaningful. Calculating the corrections for various 
differential distributions, we show that these can be substantial 
and significantly different from those within the SM. Furthermore, 
the corrections have a very distinctive flavour dependence. 
And finally, the scale dependence of the cross sections are 
greatly reduced once the NLO corrections are taken into account.}

\begin{document}

\setcounter{footnote}{0}
\renewcommand{\thefootnote}{\arabic{footnote}}

\section{Introduction}

While the Standard Model (SM) remains a very consistent explanation
for nearly all data pertaining to high energy physics experiments, a
few small discrepancies persist. Furthermore, there are theoretical
issues that cannot even be addressed within the framework of the SM
alone. Examples include the replication of the fermion families, the
naturalness problem associated with the Higgs scale, charge
quantization, the baryon asymmetry in the universe, the presence of
dark matter etc.. Clearly, an answer to such vital questions may be
obtained only in a model much more ambitious than the SM. Candidates
for the role include, amongst others, supersymmetry~\cite{susy}, grand
unification~\cite{Pati:1974yy,GUTS} (with or without supersymmetry),
family symmetries (gauged or otherwise) and compositeness for quarks
and leptons~\cite{llmodel}. In general, each such scenario (with its
peculiar strengths and weaknesses) is associated with an individual
set of tell-tale signatures. On the other hand, if the SM is to be a
valid effective low-energy description of such bigger structures, one
should be able to construct, within the ambit of the SM, operators
that would encapsulate a class of remnant effects that could pertain
to any of these scenarios. We illustrate this explicitly in the
context of one of the above-mentioned scenarios.

The replication of fermion families suggests the possibility of
quark-lepton compositeness. In such theories, the fundamental
constituents, very often termed {\em preon}s\cite{preon}, experience
an hitherto unknown force on account of an asymptotically free but
confining gauge interaction\cite{preon_binding}.  At a characteristic
scale $\Lambda$, this interaction would become very strong leading to
bound states (composites) which are to be identified as quarks and
leptons.  In most such models\cite{comp_mod, Additional_comp}, quarks
and leptons share at least some common constituents.  Since the
confining force mediates interactions between such constituents, it
stands to reason that these, in turn, would lead to interactions
between quarks and leptons that go beyond those existing within the
SM. Well below the scale $\Lambda$, such interactions would likely be
manifested through an effective four fermion contact 
interaction~\cite{cont_inter} term
that is an invariant under the SM gauge group. A convenient and
general parametrization of such interactions is given
by\cite{llmodel,Taekoon} \beq {\cal L} = {4 \pi \over \Lambda^2} \Big[
\eta_{ij} \, (\bar{q} \, \gamma^{\mu} \, P_i \, q) \,
          (\bar{l} \, \gamma_{\mu} \, P_j \, l) 
+ \xi_{ i j }  \, (\bar q \, P_i \, q) \, (\bar \ell \, P_j  \, \ell)
 \Big],
     \label{lagrangian} 
\eeq 
where $i, j = L, R$ and $P_i$ are the chirality projection operators.
Note that the Lagrangian of eqn.(\ref{lagrangian}) is by no means a
comprehensive one and similar operators involving the quarks alone (or
the leptons alone) would also exist. However, for our purpose, it
would suffice to consider only eqn.(\ref{lagrangian}). Within this
limited sphere of applicability, the strength of the interaction may
be entirely absorbed in the scale $\Lambda$, and the couplings
$\eta_{ij}$ and $\xi_{ij}$ canonically normalized to $\pm 1$.

While we have sought to motivate eqn.(\ref{lagrangian}) in the context
of compositeness, these are by no means the only scenarios ones that
can give rise to such an effective interaction lagrangian. As is well
known, a four-fermion process mediated by a particle with a mass
significantly higher than the energy transfer can be well approximated
by a contact interaction~\cite{cont_inter} term with a generic form as
in eqn.(\ref{lagrangian}). Examples include theories with extended
gauge sectors, leptoquarks~\cite{leptoquarks}, sfermion
exchange in a supersymmetric theory with broken
$R$-parity~\cite{fayet} etc..  In all such cases, on integrating out
fields with masses $M_i \gsim \Lambda$\cite{Hasenfratz:1987tk}, a
series of such higher-dimensional terms obtain. Those in
eqn.(\ref{lagrangian}) are just the lowest order (in $\Lambda^{-1}$)
ones.

Several points are in order here
\begin{itemize}
\item In general, integrating out the heavy fields would result in 
 an almost infinite number of higher-dimensional operators. The terms in 
eqn.(\ref{lagrangian}) are just some of the lowest order (in $\Lambda^{-1}$)
 ones relevant to four-fermion processes.

\item For a given model, the couplings $\eta_{ij}$ and $\xi_{ij}$
generated by the process of integrating out heavy fields would be
related to each other.  Such relations are model-specific and
determined, to a large extent, by the flavour structure of the parent
theory. As already indicated, we shall not consider any such 
flavour structure, but hold $\eta, \xi = \pm 1$. 

\item Even a requirement such as $SU(2) \otimes U(1)$ invariance for the 
    effective Lagrangian would imply a relation between such terms, but 
    involving different fields. However, we shall concern ourselves with 
    only those involving $q \bar q \ell^+ \ell^-$, noting that the results 
    would be essentially the same for its $SU(2)$ cousins.

\item Low energy observables (for example, meson decays) lead to 
severe constraints \cite{low_energy} on several of these couplings 
and even more so, on their products. Although many of these bounds 
were derived in the context of specific ultraviolet completions, it is 
easy to see that they are equally applicable to the generic contact 
interactions.

\item Note that the vector--axial vector ($VA$) or scalar--psedoscalar 
($SP$) nature of the $\eta$-- and $\xi$--couplings do not
necessarily reflect the spin of the integrated out field that led 
to such terms.
\end{itemize}

Clearly, operators such as these could, in principle, lead to 
significant phenomenological consequences in collider experiments, 
whether $e^+ e^-$~\cite{opal}, $e \, P$~\cite{HERA2} or hadronic. 
Given the higher-dimensional nature of ${\cal L}$, it is obvious that
the consequent effects would be more pronounced at higher energies. In
other words, the fractional deviation over the SM expectations would
be concentrated more at higher invariant masses $M$~\cite{M_dist},
with possibly some nontrivial dependence on the rapidity $y$ as
well~\cite{MY_dist}. For example, composite quarks and electrons have
been proposed as a possible explanation for the high-$Q^2$ anomaly at
HERA~\cite{HERA2}. Some of the best constraints on compositeness, 
for example, came from the OPAL~\cite{opal} and CDF~\cite{cdfprl} experiments.
More recently the 
measurement of the Drell-Yan cross section\cite{Drell_Yan} at high 
invariant masses 
set the most stringent limits on contact interactions 
of the type given in eqn.(\ref{lagrangian}). For example, within 
the $VA$-type interaction scenario, the scale $\Lambda$ is constrained to be 
$\Lambda \gsim 3.3$--6.1 TeV~\cite{D0_coll, NKMondal}, with the 
bound depending on the chirality structure of the operator.

As is well known, QCD corrections can alter quite significantly the 
cross sections at a hadronic collider. Thus, these may have serious 
bearing on the discovery potential of such experiments. 
Even for as simple a process as 
Drell-Yan, the leading order (LO) results seriously underestimate 
the cross sections. This has led to the incorporation of the  
next-to-leading order (NLO) or
next-to-leading log (NLL)\cite{NLL,Martin}
results in Monte Carlos codes \cite{NLL} or event
generators such as JETRAD\cite{JETRAD}. 
However, no calculations exist for the higher order QCD corrections
to cross sections mediated by a  generic contact interaction. 
Consequently, all 
extant collider studies of contact interaction have either been based on 
just the tree level calculations,  or, in some cases, have imlicitly assumed 
that the higher order corrections are exactly the same as in the SM.
Clearly, this is an unsatisfactory state of affairs and, in this paper, 
we aim to rectify this by calculating the next-to-leading order QCD
corrections for both the $VA$-type and the $SP$-type contact interactions.

It might be argued that, such theories being nonrenormalizable, any 
higher-loop calculation is fraught with danger. However, the very structure 
of such terms (namely the current--current form of the Lagrangian) along with 
the fact that only one of the currents comprises coloured fields allows 
us to reliably calculate  QCD corrections. This holds not only for 
the specific interaction in question, but also for other theories that 
satisfy the abovementioned criterion~\cite{ravi_gravi}. On the other hand, 
were we to attempt to calculate the NLO electroweak corrections, it is by no 
means certain that similar 
levels of reliability or usefulness could be reached.

The rest of the article is organised as follows. In Section 2, we 
start by outlining the general methodology and follow it up with the 
explicit calculation of the NLO corrections to the differential 
distribution in the dilepton invariant mass. In the following section, we 
consider the rapidity distributions. Section 4 contains our numerical results. 
And finally, we summarize in Section 5.
\section{NLO corrections}
        We consider lepton pair production at a hadron collider
in the context of a generic contact interaction as exemplified by 
eqn.(\ref{lagrangian}). In other words, the process is
\beq
P(p_1) + P^{\!\!\!\!\!\!^{(-)}}(p_2) \rightarrow 
l^{+}(l_1) + l^{-}(l_2) + X(p_X)
\eeq
where $p_i$ denote the momenta of the incoming hadrons and 
$l_i$ those for the outgoing leptons. 
Similarly, the inclusive
hadronic state denoted by $X$ carries momentum $p_X$. The hadronic 
cross section is defined in terms of the partonic cross sections
 convoluted with the appropriate parton distribution functions $f_a^P(x)$
and is given by 
\beq
2 S {d\sigma \over d Q^2}^{P_1 P_2} = \sum_{ab= q,\bar{q},g} \int_0^1 dx_1\: 
\int_0^1 dx_2 \: f^{P_1}_a(x_1) \: f^{P_2}_b(x_2) \, 
\int_0^1 dz ~2\, \hat{s}\; {d\sigma^{a b } \over d Q^2} \, 
\delta(\tau - z x_1 x_2)
\label{eq:hadr_cross1}
\eeq
with $x_i$ being the fraction of the initial state hadron's momentum 
carried by the parton in question. In other words, the parton momenta
$k_i$ are given by $k_i = x_i \,  p_i$. The other variables are defined 
as 
\beq
\barr{rclcrclcrcl}
S &\equiv& (p_1+p_2)^2 
& \qquad &  
\hat{s} &\equiv& (k_1+k_2)^2  
 & \qquad &  
Q^2 &\equiv&  (l_1+l_2)^2 
\\[2ex]
\tau&\equiv& \dis {Q^2 \over S}
& \qquad &  
 z &\equiv& \dis {Q^2 \over \hat{s} } 
& \qquad &  
\tau &\equiv& z\,x_1\,x_2 \ .
\earr
\eeq
It is convenient to symbolically cast the matrix element for the process 
as a sum of several current-current pieces with a ``propagator'' in between. 
In other words, 
\beq
{\cal M}^{\rm Total} = \sum_j \, {\cal J}^{\rm Had}_j 
                        \cdot  P_j \cdot {\cal J}^{\rm Lept}_j 
\eeq
where the dots ($\cdot$) denote Lorentz index contractions as appropriate 
and the propagators $P_j$ are 
\beq
\barr{rclcl c rclcl}
P_{\gamma} &=& \dis {i \over Q^2} g_{\mu \nu} 
           & \equiv & g_{\mu \nu} \tilde{P}_{\gamma}
& \qquad & 
P_Z &=& \dis {i \, g_{\mu \nu} \over Q^2 - M_Z^2 - i M_Z \, \Gamma_Z } 
& \equiv &  g_{\mu \nu} \tilde{P}_Z
\\[2ex]
P_{VA} &=& \dis {4 \pi \over \Lambda^2} & \equiv & \tilde{P}_{VA}
& \qquad & 
 P_{SP} &=& \dis {4 \pi \over \Lambda^2} & \equiv & \tilde{P}_{SP}.
\earr
\eeq
With this definition, 
the partonic cross section for the process $a(k_1) + b(k_2)
\rightarrow j(q) + \sum_{i}^m X(p_i)$ is given by
\beq
2 \hat{s} {d\sigma^{a b} \over d Q^2} = {1 \over 2 \pi} \sum_{jj^{\prime}
= \gamma, Z, VA, SP} \int d{PS}_{m+1} 
\:|{\cal M}^{ab \rightarrow jj^{\prime}}|^2 \cdot P_j(Q^2)\cdot  P_{j^{\prime}}^{*}(Q^2)
\cdot{\cal L}^{jj^{\prime} \rightarrow l\, l^{\prime} } \ ,
    \label{eq:partonic_1}
\eeq
where $|{\cal M}^{ab \rightarrow jj^{\prime}}|^2$ denotes the 
square of the hadronic current, and $dPS_{m+1}$ the ($m+1$)--body 
phase space element, viz. 
\beq
\barr{rcl}
dPS_{m+1} &=& \int \prod_i^{m}\Bigg({d^np_i \over (2\pi)^n} \: 2 \pi
\,\delta^{+}(p_i^2)\Bigg) 
\,{d^nq \over (2\pi)^n} \: 2 \pi\, \delta^{+}(q^2 - Q^2)
 \nonumber\\[2ex]
 && \times 
(2\pi)^n\, \delta^{(n)}\Big(k_1+k_2 -q -\sum_i^m p_i\Big)
\earr
\eeq
where $n$ is the dimension of spacetime and $\delta^+(x)$ carries its 
usual meaning.  The leptonic tensor, given by
\beq
{\cal L}^{jj^{\prime} \rightarrow \,l\, l^{\prime} } =
\int \prod_i^{2}\Bigg({d^nl_i \over (2\pi)^n} \: 2 \pi
\,\delta^{+}(l_i^2)\Bigg) 
(2\pi)^n\, \delta^{(n)}\Big(q - l_1 - l_2\Big)
|{\cal M}^{jj^{\prime}\rightarrow \,l^+ l^-}|^2 \ ,
\eeq
is straightforward to compute and leads to
 \beq
 {\cal L}_{jj^{\prime} \rightarrow l\, l^{\prime} } 
    = \left\{ 
       \barr{rcl}
       \dis\Big( -g_{\mu \nu} + {q_{\mu} q_{\nu} \over Q^2} \Big) \, 
                 {\cal L}_{jj^{\prime}}(Q^2)  & \qquad & 
                    (j, j^{\prime} = \gamma,  Z, VA) \\[1ex]
               {\cal L}_{SP}(Q^2) & & (j=j^{\prime} = SP)
	\earr
     \right.
\eeq
 with
 \beq
\barr{rcl c rcl}
 {\cal L}_{\gamma \gamma}(Q^2) &=& \dis {2 \alpha \over 3} \, Q^2,
& \qquad & 
 {\cal L}_{Z Z}(Q^2) & = & \dis {\alpha \over 3} 
  \Bigg(\Big(g^R_l\Big)^2 + \Big(g^L_l\Big)^2 \Bigg) Q^2
\\[2ex]
 {\cal L}_{\gamma Z}(Q^2) &=& \dis 
              {- \alpha \, g_\ell^V \over 6 } \, Q^2,
& &  {\cal L}_{SP}(Q^2) & =  & Q^2
\earr
 \eeq
In the above, $\alpha$ denotes the fine structure constant, 
while $g_a^{L, R}$ parametrize
the couplings of the left-- and right-chiral fermionic fields to the $Z$, viz.
\beq
g_a^V = \frac{1}{2} \left(g_a^R +g_a^L \right) \ , \qquad
  g_a^L = - \, e_a \, \tan \theta_W  \ ,\qquad
g_a^R = - 2 \, T_a^3 \, \csc 2\theta_W - \, e_a \, \tan \theta_W  
\eeq
in terms of the Weinberg angle ($\theta_W$)  and the electric 
charge ($e_a$) of the fermion in question.

On substituting for  ${\cal L}_{jj^{\prime} \rightarrow l\, l^{\prime} }$ 
in eqn.(\ref{eq:partonic_1}), we have, for the hadronic cross section,
\beq
2 S {d\sigma^{P_1 P_2} \over dQ^2}(\tau, Q^2) = {1 \over 2 \pi}
\sum_{j,j^{\prime}= \gamma, Z, VA, SP}
\tilde{P}_j(Q^2) \: \tilde{P}_{j^{\prime}}^{*}(Q^2) \:{\cal L}_{jj^{\prime}}(Q^2)
\: W_{jj^{\prime}}^{P_1 P_2}(\tau, Q^2)
\eeq
where the hadronic structure function $W$ is defined to be
\beq
W_{jj^{\prime}}^{P_1 P_2}(\tau, Q^2) = \sum_{a,b,j,j^{\prime}} \int_0^1 dx_1\: 
\int_0^1 dx_2 \: f^{P_1}_a(x_1) \: f^{P_2}_b(x_2) 
\int_0^1 dz ~ \delta(\tau - z x_1 x_2) 
\bar{\Delta}^{jj^{\prime}}_{ab}(z, Q^2, \epsilon) \ .
\eeq
All that remains is to calculate the bare partonic coefficient function
$\bar \Delta$: 
\beq
\bar{\Delta}^{jj^{\prime}}_{ab}(z, Q^2, \epsilon) = 
\int dPS_{m+1} 
\:|{\cal M}^{ab \rightarrow jj^{\prime}}|^2 \,T_{jj^{\prime}}(q).
\eeq
where $T_{jj}$ depends upon the spin of the current in question, viz
\beq
\barr{rcl}
T_{jj^{\prime}}(q) &=& \dis \Big( -g_{\mu \nu} + 
{q_{\mu} q_{\nu} \over Q^2} \Big)
~~~~~~~~~ (j, j^{\prime} = \gamma \ , ZZ \ , VA)
\\[1.5ex]
T_{SP}(q) &=& 1  
\earr
\eeq
To compute the $Q^2$ distribution of the dilepton pair, one then 
has to calculate the square of the hadronic matrix element $|{\cal M}^{ab
\rightarrow jj^{\prime}}|^2 \: T_{jj^{\prime}}(q)$, preferably 
in a  suitable frame so as to render the integrations over the 
phase space and $z$ easy. 

Note that, the bare partonic coefficient function
$\bar \Delta$ is a singular object, suffering from each of 
ultraviolet, soft and collinear divergences. To handle these, we adopt 
dimensional regularisation. The renormalization procedure for 
the $VA$-type interactions is quite established and may be found, for example, 
in Ref.\cite{ravi_gravi}. Note that, for the
$SP$ type interaction, one-loop corrections results in an extra 
term---proportional to $\ln\Big(Q^2 / \mu^2\Big)$---as compared to the 
$VA$ interactions~\cite{1st_paper}. This, of course, is not unexpected, 
as contrary to the usual conserved vector currents, a scalar current is 
renormalized by QCD interactions. It is easy to see that this extra 
term is precisely the one that is absorbed into the bare contact 
interaction coupling constant in defining the renormalized coupling 
$\xi_{ij}$. 

To the ultraviolet regularized (and renormalized) operator, we 
must add the contribution from the real emission diagrams (gluon 
bremsstrahlung as well as the Compton process), and this exercise 
leaves us with a quantity that suffers only from collinear singularities.
The latter, of course, can be removed through mass factorization. 
If $\mu_F$ be the  factorization scale, then  Drell-Yan coefficient 
functions, after mass factorization by $\Delta^{jj^{\prime}}_{ab}$,  
are related to the bare functions through 
\beq
 \bar{\Delta}^{jj^{\prime}}_{ab}(z, Q^2, \epsilon) =  \sum_{c,d} 
\Gamma_{ca}(z,\mu_F^2, 1/\epsilon) \otimes \Gamma_{db}(z,\mu_F^2, 1/\epsilon)
\otimes \Delta^{jj^{\prime}}_{cd}(z, Q^2, \mu_F^2)
\label{eq:mass_fact_coeff_func}
\eeq
with the convolution defined to be 
\beq
f \otimes g (x) = \int_x^1 \: {d y \over y} \: f(y) \: g({x \over y}) \ .
\eeq
The kernels $\Gamma_{ab}$ are related to the leading order 
Altarelli-Parisi splitting
functions\cite{Alteralli_splitting} $P^{(0)}_{ab}(z)$ through 
\beq
\barr{rcl}
\Gamma_{ab}(z,\mu_F^2, 1/\epsilon) &=& \dis \delta_{ab} \: \delta(1-z)
+ {\alpha_s(\mu^2) \over 4 \, \pi \, \epsilon }  \; 
\Gamma_{ab}^{(1)}(z,\mu_F^2)
\\[2ex]
 &=& \dis
 \delta_{ab} \: \delta(1-z) + {a_s  \over \epsilon} 
\Bigg({\mu_F^2 \over \mu^2} \Bigg)^{\epsilon/ 2} P^{(0)}_{ab}(z)
\\[2ex]
a_s & = & \dis {\alpha_s(\mu^2) \over 4 \pi}
\earr
\eeq
Expanding eqn.(\ref{eq:mass_fact_coeff_func}) to order $a_s$
we have
\beq
\barr{rcl}
\bar{\Delta}^{jj^{\prime}}_{q \bar{q}} &=& \dis
\Delta^{(0),{jj^{\prime}}}_{q \bar{q}}  + a_s {2 \over \epsilon}
\Gamma_{q \bar{q}}^{(1)} \otimes \Delta^{(0),{jj^{\prime}}}_{q \bar{q}}
+ a_s \Delta^{(1),{jj^{\prime}}}_{q \bar{q}}
\\[2ex]
\bar{\Delta}^{jj^{\prime}}_{q g} &=& \dis a_s {2 \over \epsilon}
\Gamma_{q g}^{(1)} \otimes \Delta^{(0),{jj^{\prime}}}_{q g}
+ a_s \Delta^{(1),{jj^{\prime}}}_{q g}
\earr
\eeq
thereby leading us to the finite coefficient functions. 
The physical hadronic cross section may be obtained by folding these
finite coefficient functions with appropriate parton distribution
functions. For the sake of completeness, we present the results below.
To begin with, we denote the renormalized parton-parton fluxes by 
$H_{ab} (x_1, x_2, \mu_F^2)$ where 
\beq
\barr{rcl}
H_{q\bar{q}}(x_1, x_2, \mu_F^2) &=& \dis f^{P_1}_q(x_1,\mu_F^2) f^{P_2}_{\bar{q}}(x_2,\mu_F^2)
+ f^{P_1}_{\bar{q}}(x_1,\mu_F^2) f^{P_2}_q(x_2,\mu_F^2)
\\[2ex]
H_{g q}(x_1, x_2, \mu_F^2) &=& \dis f^{P_1}_g(x_1,\mu_F^2)\Big(f^{P_2}_q(x_2,\mu_F^2)
+ f^{P_2}_{\bar{q}}(x_2,\mu_F^2) \Big)
\\[2ex]
H_{q g}(x_1, x_2, \mu_F^2) &=& \dis H_{g q}(x_2, x_1, \mu_F^2) \ .
\earr
\eeq
Then, the inclusive differential cross section may be expressed as 
\beq
\barr{rcl}
\dis 2 S {d\sigma^{P_1 P_2} \over dQ^2}(\tau, Q^2) &=& \dis
\sum_q \int_0^1 dx_1\: 
\int_0^1 dx_2 
\int_0^1 dz ~ \delta(\tau - z x_1 x_2) \\[1.5ex]
& & \dis \hspace*{6ex}  \left[  {\cal F}_{SM+VA,q} \; {\cal G}_{SM + VA, q} 
         + {\cal F}_{SP,q} \; {\cal G}_{SP, q} \right]
\\[3ex]
{\cal G}_{SM + VA, q} & \equiv & 
H_{q\bar{q}}(x_1,x_2,\mu_F^2)\Big\{\Delta^{(0),SM}_{q \bar{q}}(z,Q^2,\mu_F^2)
+ a_s \Delta^{(1),SM}_{q \bar{q}}(z,Q^2,\mu_F^2) \Big\} 
\\[2ex]
&+ &
\Big\{H_{q g}(x_1,x_2,\mu_F^2) + H_{gq}(x_1,x_2,\mu_F^2)\Big\} a_s
\Delta^{(1),SM}_{q g}(z,\mu_F^2)
\\[3ex]
{\cal G}_{SP, q}&\equiv& \dis
H_{q\bar{q}}(x_1,x_2,\mu_F^2)\Big\{\Delta^{(0),SP}_{q \bar{q}}(z,Q^2,\mu_F^2)
+ a_s \Delta^{(1),SP}_{q \bar{q}}(z,Q^2,\mu_F^2) \Big\}
\\[2ex]
&+ &
 \Big\{H_{q g}(x_1,x_2,\mu_F^2) + H_{g q}(x_1,x_2,\mu_F^2)\Big\} a_s
\Delta^{(1),SP}_{q g}(z,\mu_F^2)
\label{dsig:dm}
\earr
\eeq
with the constants ${\cal F}_{SM+VA,q}$ and ${\cal F}_{q}^{SP}$ 
containing all the dependences on the coupling constants and 
propagators, namely, 
\beq
\barr{rcl}
{\cal F}_{SM+VA,q} &=& \dis {4 \alpha^2 \over 3 }
  \Bigg[\Bigg\{ \frac{e^2_q}{Q^2}
- 2 \, e_q \, g_l^V \, g_q^V \, Z_Q \, \frac{Q^2 - M_Z^2}{Q^2}
\\[2ex]
& & \dis \hspace*{6ex} + 
 \frac{1}{4}\, 
   \Big( (g_l^R )^2 + (g_l^L )^2 \Big) \Big( (g_q^R )^2 + (g_q^L )^2 \Big)
\, Z_Q  
\Bigg\}
\\  [2ex]
&& \dis \hspace*{4ex} + 
    {2 \over \alpha \Lambda^2} \, 
     \Bigg\{ - {e_q } \sum_{i, j = L, R} \eta_{i j}
+ {Z_Q (Q^2 - M_Z^2)} \, 
    \sum_{i, j = L, R} \eta_{i j}  g_q^i g_l^j \Bigg\}
\\[2ex]
&& \dis \hspace*{4ex} + 
 {Q^2 \over \alpha^2 \Lambda^4} \sum_{i, j = L, R} |\eta_{ij}|^2 
\Bigg]
\\[3ex]
{\cal F}_{SP, q} &=& \dis {Q^2\over \Lambda^4}
    \,  \sum_{i, j = L, R} |\xi_{ij}|^2 
\\[3ex]
Z_Q & \equiv & \dis \frac{Q^2}{ (Q^2 - M_Z^2)^2 + \Gamma_Z^2 \, M^2_Z}
\earr
   \label{eq:consts_F}
\eeq

For the vector-axial vector couplings, the results for the 
coefficient functions are analogous to the case of the 
SM~\cite{ravi_gravi}, namely
\beq
\barr{rcl}
\Delta_{q\bar{q}}^{(0),VA} &=& \dis{ 2 \pi \over N} \, \delta(1-z)
\earr
\eeq
\beq
\barr{rcl}
\Delta^{(1),VA}_{q\bar{q}} &=& \dis {8 \, \pi \,  C_F \over N} 
\Bigg[
\Big\{-4 + 2 \zeta(2) \Big\}\, \delta(1-z) 
-  \, (1+z) \, \ln {(1-z)^2 \over z } - \,2 \, {\ln(z) \over 1-z} +
\\[2ex]
& &  
\left\{ {2 \over (1-z)_{+}} + {3 \over 2} \, \delta(1-z) - (1+z) 
\right\} \, \ln\Bigg({Q^2 \over \mu_F^2}\Bigg)
 + \,4 \Bigg( {\ln(1-z) \over 1-z} \Bigg)_{+}
\Bigg] 
\\[3ex]
\Delta^{(1),VA}_{q (\bar{q}) g} &=& \dis {2 \pi \over N} \,T_F
\Bigg[
2\,\Big\{1 - 2\,z + 2\,z^2\Big\} \,\ln\Bigg({Q^2 (1-z)^2 \over z \mu_F^2}\Bigg)
+1 + 6\, z - 7\,z^2
\Bigg] \ .
\earr
\eeq
For the scalar-pseudoscalar
couplings, on the other hand, the LO coefficient function is given by
\beq
\Delta_{q\bar{q}}^{(0),SP} = { 2 \pi \over N} \, \delta(1-z)
\eeq
while at the next-to-leading order coefficient functions are
\beq
\barr{rcl}
\Delta^{(1),SP}_{q\bar{q}} &=& \dis { 4 \,  \pi \, C_F \over N} 
\Bigg[
\Big\{-2 + 4 \zeta(2) \Big\}\, \delta(1-z) + 2\,(1-z)
+ \,4 (1+z^2) \Bigg( {\ln(1-z) \over 1-z} \Bigg)_{+}
\nonumber\\[2ex]
&&\dis \hspace*{4em}
+2 \, {1+z^2 \over (1-z)_{+}} \ln\Bigg({Q^2 \over z \mu_F^2}\Bigg)
+ 3\,\delta(1-z) \, \ln\Bigg({Q^2 \over \mu_F^2} \Bigg)
\Bigg],  
\\[2ex]
\Delta^{(1),SP}_{q (\bar{q}) g} &=& \dis {2 \, \pi \, T_F \over N} 
\Bigg[
 2\,\Big(1 - 2\,z + 2\,z^2\Big) \,\ln\Bigg({Q^2 (1-z)^2 \over z \mu_F^2}\Bigg)
+ (1-z)(7 z -3)
\Bigg].
\earr
\eeq
The $SU(N)$ color factors in the above equations are
\beq
C_F = {N^2-1 \over 2 N}, \hspace*{1cm}
C_A = N, \hspace*{1cm} T_F = {1 \over 2}.
\eeq
\section{Differential cross sections with respect to dilepton
rapidity} \label{sec:Y_dist} 

Having considered, in the previous
section, the differential distributions with respect to the dilepton invariant
mass, we now consider a second variable of interest, namely the rapidity 
of the pair. The latter can be expressed as 
\beq
Y = {1 \over 2} \log\Bigg({p_2\cdot q \over p_1 \cdot q} \Bigg) 
  =
{1 \over 2} \log\Bigg({q^0_{CMH} - q^3_{CMH}
\over q^0_{CMH} + q^3_{CMH}} \Bigg)
\eeq
with the second equality valid in the center of mass frame of the hadrons.
Thus, the rapidity distribution may be computed simply by introducing
the identity 
\[ 
\int dY \delta\Bigg(Y - {1 \over 2}
\log\Bigg({p_2\cdot q \over p_1 \cdot q} \Bigg)\Bigg) = 1,
\]
in eqn.(\ref{eq:hadr_cross1}). This leads to 
\beq
2 S {d\sigma^{P_1 P_2} \over
dQ^2 dY}(\tau, Y, Q^2) = {1 \over 2 \pi} \sum_{j,j^{\prime}= \gamma,
Z, VA, SP} \tilde{P}_j(Q^2) \, \tilde{P}_{j^{\prime}}^{*}(Q^2) \, {\cal
L}_{jj^{\prime}}(Q^2) \, 
  {d\,W_{jj^{\prime}}^{P_1 P_2} \over d Y}(\tau, Y, Q^2).
\label{eq:hadron_cross_Y}
\eeq
where the hadronic structure functions are given by
\beq
\barr{rcl}
\dis {d\,W_{jj^{\prime}}^{P_1 P_2} \over d Y}(\tau, Y, Q^2) &=&
\dis \sum_{a,b,j,j^{\prime}} \int_0^1 dx_1\: 
\int_0^1 dx_2 \: f^{P_1}_a(x_1) \: f^{P_2}_b(x_2) \;
\int_0^1 dz \, \delta(\tau - z x_1 x_2) 
\\[2ex]
&&\dis \hspace*{1em}\times \, \int d{PS}_{m+1} 
\:|{\cal M}^{ab \rightarrow jj^{\prime}}|^2 \, T_{jj^{\prime}}(q) \, 
\delta\Bigg(Y - {1 \over 2} \log\Bigg({p_2\cdot q \over p_1 \cdot q} 
\Bigg)\Bigg)
\label{eq:hadron_str_fun_Y}
\earr
\eeq
We start with the leading order case which involves just the calculation of
the square of the matrix element for the process 
$a(k_1) + b(k_2) \rightarrow j(q)$. The relevant phase space element
corresponds to that for a $(0+1)$-body final state, and
\beq
\barr{l}
\dis \int dPS_{0+1} \int dz \delta\Bigg(Y - {1 \over 2} \log\Bigg({p_2\cdot q \over p_1 \cdot q} 
\Bigg)
\Bigg) \delta(\tau - z x_1 x_2) 
\\[2ex]
\dis \hspace*{5em} 
= {2 \pi \over Q^2} \int dz \, 
\delta\Bigg(Y - {1 \over 2} \,  \log \, {x_1 \over x_2} \, \Bigg) \;
\delta(1-z) \  \delta(\tau - z x_1 x_2) \ .
\earr
\label{eq:Y_constraint}
\eeq
The integration over the rest of the variables is simplified, particularly 
in the context of the NLO corrections, by effecting a change of variables, 
namely 
\beq
(Y, \tau) \longrightarrow (x_1^0, x_2^0) \equiv 
\left(\sqrt{\tau}\, e^{Y}, \sqrt{\tau}\, e^{-Y} \right)
\eeq
Then it follows that 
\beq
\barr{rcl}
\dis
{2 \pi \over Q^2} \int dz
\delta\Bigg(Y - {1 \over 2} \log\Bigg({x_1 \over x_2} 
\Bigg)\Bigg) \, 
 \delta(1-z) \, \delta(\tau - z x_1 x_2)  
\; |{\cal M}^{ab \rightarrow jj^{\prime}}|^2 \, T_{jj^{\prime}}
\\[2ex]
\dis \hspace*{3em} 
= {2 \pi \over Q^2} \,  \delta(x_1-x_1^0) \, \delta(x_2-x_2^0) \; 
  \Bigg[ \, |{\cal M}^{ab \rightarrow jj^{\prime}}|^2 \, T_{jj^{\prime}}
  \Bigg]_{z \, = \, 1} \ ,
\earr
\eeq
rendering the remaining integrals trivial and thereby giving us the Born-level 
result for the $Y$-distribution. 
Having set the formalism, we 
may now calculate the next-to-leading-order contribution to the same. 
This involves the computation of matrix element squared
for the processes $a(k_1) + b(k_2) \rightarrow j(q) + c(k)$. The 
$(1+1)$-body phase space integration can be performed in the 
CM frame of the incoming partons wherein the particle momenta may 
be parametrised as
\[
\barr{rcl}
k_1 &=& \dis {\sqrt{\hat{s}} \over 2} (1,0,\cdots,0,1)
\\
k_2 & = & \dis {\sqrt{\hat{s}} \over 2} (1,0,\cdots,0,-1)
\\[2ex]
-q &=& \dis 
{\sqrt{\hat{s}} \over 2} (1+z,0,\cdots,-(1-z)\sin\theta,-(1-z)\cos\theta)
\\[2ex]
-k &=& \dis
{\sqrt{\hat{s}} \over 2} (1-z,0,\cdots,(1-z)\sin\theta,(1-z)\cos\theta)
\earr
\]
Writing $\cos\theta = 2 y - 1$, the two delta functions reduce to
\beq
\barr{rcl}
\dis
\delta\Bigg(Y - {1 \over 2} \log \: {p_2\cdot q \over p_1 \cdot q}\Bigg) 
& = & \dis
\delta\Bigg(Y - {1 \over 2} \log \, {x_1\big(1-y(1-z)\big) \over
x_2\big(z+y(1-z)\big)}\Bigg)
\\[3ex]
& = & \dis
{2 x_1 x_2 x_1^0 x_2^0 (x_1 x_2 + x_1^0 x_2^0) \over
(x_1 x_2 - x_1^0 x_2^0) (x_1 x_2^0 + x_1^0 x_2)^2}\, \delta(y - y^{*})
\\[3ex]
\delta(\tau - z x_1 x_2) &=& \dis {1 \over x_1 \, x_2} \, \delta(z - z^{*})
\label{del_constraint}
\earr
\eeq
where,
\begin{eqnarray}
y^* = {x_2 x_2^0 (x_1 + x_1^0)(x_1 - x_1^0) \over
(x_1 x_2 - x_1^0 x_2^0) (x_1 x_2^0 + x_1^0 x_2) },
\hspace{1.5cm}
z^* = {x_1^0 x_2^0 \over x_1 x_2}
\label{yz_constraint}
\end{eqnarray}
The above relations can be used to obtain
\beq
\barr{l}
\hspace*{-3em}
\dis 
\int dPS_{1+1} \int dz \, \delta\Bigg(Y - {1 \over 2} \,
                   \log \, {p_2\cdot q \over p_1 \cdot q} 
                             \Bigg) \; 
\delta(\tau - z x_1 x_2) \;
|{\cal M}^{ab \rightarrow jj^{\prime}}|^2 \; T_{jj^{\prime}}
\\[2ex]
\dis
= {1 \over 8 \pi} \, 
   \Bigg({Q^2 \over 4 \pi}\Bigg)^{\epsilon / 2} \, 
   {1 \over \Gamma(1+\epsilon/2)} \; 
{2 x_1^0 x_2^0 (x_1 x_2 + x_1^0 x_2^0) \over x_1 x_2 (x_1 x_2^0 + x_1^0 x_2)^2}
\; 
\left[ \, 
  |{\cal M}^{ab \rightarrow jj^{\prime}}|^2 T_{jj^{\prime}} \right]_{y= y^*, z = z^*}
\\[2ex] 
\dis \hspace*{2em} \times
\Bigg(\frac{(x_1-x_1^0)(x_2-x_2^0)(x_1+x_1^0)(x_2+x_2^0)}
           {(x_1 x_2^0 + x_1^0 x_2)^2}  \Bigg)^{\epsilon/2}

\label{phsp_constraint}
\earr
\eeq
To obtain the  contribution to the $Y$ distribution from real 
gluon emissions, 
we substitute eqn.(\ref{phsp_constraint}) in eqn.(\ref{eq:hadron_cross_Y}).
Similarly, the virtual corrections can be obtained 
using eqn.(\ref{eq:Y_constraint}) with oneloop corrected matrix elements. 
The soft singularities cancel after adding the real emission 
contributions and virtual corrections to the 
Born process. The remaining collinear
divergences are removed by mass factorization, or, in other words, 
by replacing the bare parton distribution
with the renormalized ones using the Alteralli-Parisi kernels as follows
\begin{eqnarray}
f_a^P(z) = \sum_b \Gamma^{-1}_{ab} \otimes f_b^P(z, \mu_F^2),
\end{eqnarray}
which implies
\beq
\barr{rcl}
f_q^P(z) &=& \dis 
f_q^P(z,\mu_F^2) - {a_s  \over \epsilon} \, 
   \left[  \Gamma^{(1)}_{qq} \otimes f_q^P(z, \mu_F^2) 
         + \Gamma^{(1)}_{qg} \otimes f_g^P(z, \mu_F^2) \right] 
\\[2ex]
f_{\bar{q}}^P(z) &=& \dis 
f_{\bar{q}}^P(z,\mu_F^2) - {a_s \over \epsilon} \,
   \left[   \Gamma^{(1)}_{\bar{q}\bar{q}} \otimes f_{\bar{q}}^P(z, \mu_F^2) 
          + \Gamma^{(1)}_{\bar{q}g} \otimes f_g^P(z, \mu_F^2) \right] 
\\[2ex]
f_{g}^P(z) &=& \dis 
   f_{g}^P(z,\mu_F^2) - {a_s \, n_f \over \epsilon} \, 
 \Big[  \Gamma^{(1)}_{g q} \otimes f_q^P(z, \mu_F^2) 
       + \Gamma^{(1)}_{g\bar{q}} \otimes f_{\bar{q}}(z,\mu_F^2)
  \\[2ex]
&&       +  \Gamma^{(1)}_{g g} \otimes f_g^P(z, \mu_F^2) \Big] . 
\label{mass_fact}
\earr
\eeq
Thus, we finally have, for the one-loop corrected distributions in the 
dilepton pair rapidity, 
\beq
\barr{rcl}
\dis 
2 S {d\sigma \over dQ^2 dY}(\tau, Y, Q^2) &=& \dis 
\sum_{i=q} {\cal F}_{SM+VA,q} \Bigg[ D_{q\bar{q}}^{SM}(x_1^0,x_2^0,\mu_F^2)
\\[2ex]
&& \dis \hspace*{6em}
+ D_{qg}^{SM}(x_1^0,x_2^0,\mu_F^2) + D_{gq}^{SM}(x_1^0,x_2^0,\mu_F^2)
\Bigg]
\\[2ex]
&+ &  \dis
\sum_{i=q}
{\cal F}_{SP, q}\, \Bigg[ D_{q\bar{q}}^{SP}(x_1^0,x_2^0,\mu_F^2)
\\[2ex]
&& \dis \hspace*{6em}
+ D_{qg}^{SP}(x_1^0,x_2^0,\mu_F^2) + D_{gq}^{SP}(x_1^0,x_2^0,\mu_F^2)
\Bigg]
\earr
\eeq
with the constants ${\cal F}_{SM+VA,q}$ and ${\cal F}_{SP, q}$ as 
in eqn.(\ref{eq:consts_F}). The functions $D$ can be split conveniently 
into the Born-approximation piece and the NLO corrections, viz
\beq
D_{ab}^{\eta}(x_1^0,x_2^0,\mu_F^2) = D_{ab}^{\eta,(0)}(x_1^0,x_2^0,\mu_F^2)
+ a_s D_{ab}^{\eta,(1)}(x_1^0,x_2^0,\mu_F^2) \hspace{0.5cm} 
(\eta = SM, VA, SP).
\eeq
Once again, the analytical expressions 
for the $VA$-type contact interactions are the same as those 
obtained within the SM and can be found in Ref.~\cite{ravi_gravi}. As for the 
$SP$-type interactions, while the leading-order expression is simple 
\beq
D_{q\bar{q}}^{SP,(0)}(x_1^0,x_2^0,\mu_F^2) = {\pi \over N} 
H_{q\bar{q}}(x_1^0,x_2^0,\mu_F^2)
\eeq
the NLO results are more complicated. Defining, for convenience, 
certain constants 
\beq
\barr{rcl c rcl}
{\cal \kappa}_{a_1} & = & \dis  
    \ln \, {2 \, Q^2 \, (1-x_2^0) \, (x_1-x_1^0) \over 
               \mu_F^2 \, (x_1+x_1^0) \, x_2^0} 
& \qquad & 
{\cal \kappa}_{b_1} & = & \dis
 \ln \, { Q^2 \, (1-x_2^0) \, (x_1-x_1^0) \over \mu_F^2 \, x_1^0 \, x_2^0} 
\\[2ex]
{\cal \kappa}_{c_1} &=& \dis 
 \ln \, {2 \, x_1^0 \over x_1 + x_1^0} 
& & 
{\cal \kappa}_{12} & = & \dis
\ln \, { (1-x_1^0) \, (1-x_2^0) \over x_1^0 \, x_2^0} 
\earr
\eeq
we have 
\beq
\barr{rcl}
\dis
D_{q\bar{q}}^{SP,(1)}(x_1^0,x_2^0,\mu_F^2) &=& \dis
\left( \frac{2 \, \pi \, C_F}{N} \right) \; \Bigg\{
\varphi^{q \bar q}_0 + 
    \int dx_1 \,  \varphi^{q \bar q}_1
 + 
 \int dx_1 dx_2 \, \varphi^{q \bar q}_2 
 \Bigg\}
 + \Big(1 \leftrightarrow 2 \Big)
\\[3ex]
\dis \varphi^{q \bar q}_0  & = & \dis
{1 \over 2} H_{q\bar{q}}(x_1^0,x_2^0,\mu_F^2) \, 
     \Bigg(-2 +\kappa_{12}^2 + 6 \, \zeta(2) + (3 + 2\, \kappa_{12}) 
\ln \, {Q^2 \over \mu^2_F} \Bigg)
\\[2ex]
\dis \varphi^{q \bar q}_1 & = & \dis
\frac{2 \, \kappa_{b_1} }{ x_1-x_1^0} \, H_{q\bar{q},1}(x_1,x_2^0,\mu_F^2)
\\[2ex]
& + & \dis
H_{q\bar{q}}(x_1,x_2^0,\mu_F^2) \Bigg({1-\kappa_{a_1} \over x_1}
+ {2 \kappa_{c_1} \over x_1-x_1^0} - {1+\kappa_{a_1} \over x_1^2} x_1^0\Bigg)
\\[3ex]
\dis \varphi^{q \bar q}_2 & = & \dis
{H_{q\bar{q},12}(x_1,x_2,\mu_F^2) \over (x_1-x_1^0) (x_2-x_2^0)}
\; - \; {x_2+x_2^0 \over (x_1-x_1^0) \,  x_2^2} \, H_{q\bar{q},1}(x_1,x_2,\mu_F^2)
\\[3ex]
& + & \dis
{H_{q\bar{q}}(x_1,x_2,\mu_F^2) \over 
        2 \, x_1^2 \, x_2^2 } \, 
\Bigg((x_1+x_1^0) \, (x_2+x_2^0)
+
 {x_1^2 x_2^2 + x_1^{0^2} \, x_2^{0^2} \over \, (x_1+x_1^0) \, (x_2+x_2^0)}
               \Bigg)
\earr
\eeq
and
\beq
\barr{rcl}
D_{gq}^{SP,(1)}(x_1^0,x_2^0,\mu_F^2) &=& \dis 
{2 \, \pi \, T_f\over N} \, \int {dx_1 \over x_1^3} 
\nonumber\\[2ex]
&\times&\Bigg[
 \, \varphi^{g \bar q}_1 
+ \int dx_2  \, 
  \left\{ \varphi^{g \bar q}_2 - \;  
{\varphi^{g \bar q}_3 \; H_{gq}(x_1,x_2,\mu_F^2) \over
x_2^2 \, (x_2+x_2^0) \, (x_1 x_2^0 + x_2 x_1^0)^3} 
\right\}
\Bigg]
\nonumber\\[2ex]
\dis \varphi^{g \bar q}_1  & = & \dis 
H_{gq}(x_1,x_2^0,\mu_F^2)\Bigg(2 x_1^0 (x_1-x_1^0)
+ \kappa_{a_1} \Big(x_1^{0^2} + (x_1-x_1^0)^2\Big)\Bigg)
\nonumber\\[2ex]
\dis \varphi^{g \bar q}_2  & = & \dis 
{H_{gq,2}(x_1,x_2,\mu_F^2) \over x_2-x_2^0 } \, 
 \Big(x_1^{0^2} + (x_1-x_1^0)^2\Big)
\nonumber\\[2ex]
\dis \varphi^{g \bar q}_3  & = & \dis 
- x_1^5 x_2^2 x_2^{0^3} + x_1^4 x_1^0 x_2^2 x_2^{0^2} (3 x_2 + 4 x_2^0)
\nonumber\\[2ex]
&& \dis + x_1^3 x_1^{0^2} x_2 x_2^0 (3 x_2^3 + 2 x_2^{0^3}) + 2 x_1^{0^5} x_2^2
(x_2^3 + 2 x_2^2 x_2^0 + 2 x_2 x_2^{0^2} + 2x_2^{0^3}) 
\nonumber\\[2ex]
&& \dis 
+ 2 x_1 x_1^{0^4} x_2
(-x_2^4 + x_2^3 x_2^0 + 4 x_2^2 x_2^{0^2} + 2 x_2 x_2^{0^3} + 2 x_2^{0^4})
\nonumber\\[2ex]
&& \dis
+ x_1^2 x_1^{0^3} (x_2^5 - 4 x_2^4 x_2^0 -4 x_2^3 x_2^{0^2} + 
2 x_2^2 x_2^{0^3} + 2 x_2 x_2^{0^4} + 2 x_2^{0^5}) 
\earr
\eeq
with
\beq
D_{qg}^{SP,1}(x_1^0,x_2^0,\mu_F^2) = 
D_{gq}^{SP,1}(x_1^0,x_2^0,\mu_F^2)|_{(1 \leftrightarrow 2)} \ 
\eeq
and we have used the following notations,
\beq
\barr{rcl}
H_{ab,12}(x_1,x_2,\mu_F^2) &=& H_{ab}(x_1,x_2,\mu_F^2) - H_{ab}(x_1^0,x_2,\mu_F^2)
			   -H_{ab}(x_1,x_2^0,\mu_F^2)
\nonumber\\[1ex]
&&+H_{ab}(x_1^0,x_2^0,\mu_F^2)
\nonumber\\[2ex]
H_{ab,1}(x_1,x_2,\mu_F^2) &=& H_{ab}(x_1,x_2,\mu_F^2) - H_{ab}(x_1^0,x_2,\mu_F^2)
\nonumber\\[2ex]
H_{ab,2}(x_1,x_2,\mu_F^2) &=& H_{ab}(x_1,x_2,\mu_F^2) - H_{ab}(x_1,x_2^0,\mu_F^2).
\earr
\eeq
\section{Results and Discussion}
We now present numerical results relevant for the Run II of the Tevatron
($\sqrt{S}=1.96~{\rm TeV}$) as well as for the 
LHC ($\sqrt{S}=14~{\rm TeV}$). Although our goal, namely 
the calculation of the NLO QCD corrections, would be 
quite independent of the value of the contact interaction 
scale $\Lambda$, for definiteness we choose $\Lambda = 6 \, (20) \tev$
for the Tevatron (LHC). Furthermore, in presenting our results, 
we shall consider only one of the couplings $\eta^q_{AB}$ 
and $\xi^q_{AB}$ to be non-zero and of unit strength. 

For the sake of convenience, we parametrize the cross section as 
\begin{equation}
\begin{array}{rclcl}
\sigma & = & \sigma_{\rm SM} + \sigma_{\rm intf} + \sigma_{\eta^2} 
       & \qquad & {\rm (for \ the \ VA \ case)} 
\\[2ex]
\sigma & = & \sigma_{\rm SM}  + \sigma_{\xi^2} 
       & \qquad & {\rm (for \ the \ SP \ case)} 
\end{array}
\end{equation}
and similarly for the differential cross sections. This has the 
advantage in that the total cross sections, for an arbitrary 
value of $\Lambda$ can be easily reconstructed. 

\subsection{The invariant mass distribution for the dilepton pair}

\begin{figure}[!h]
\centerline{
\epsfxsize=8.5cm\epsfysize=7.0cm
                     \epsfbox{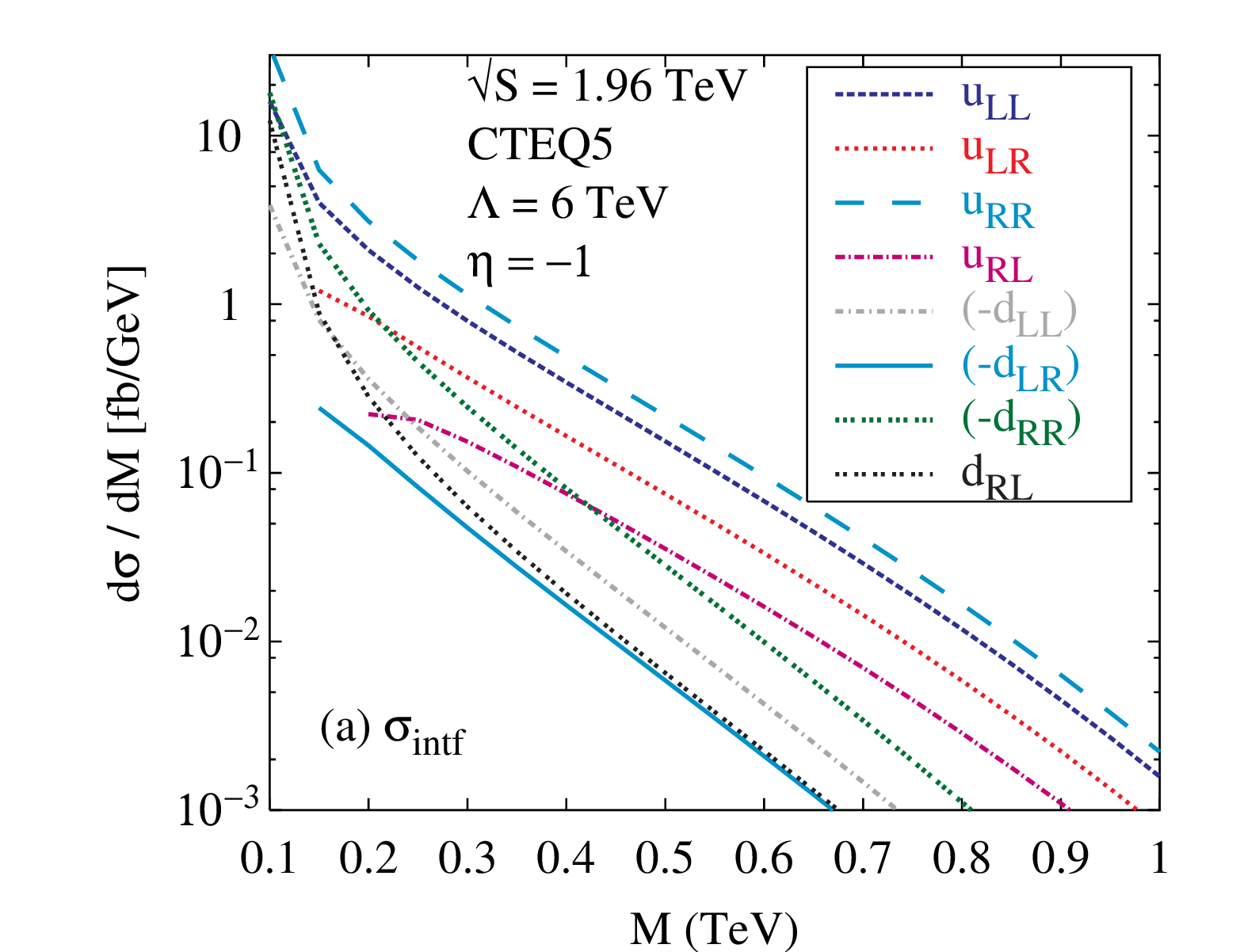}
\hspace*{-1em}
\epsfxsize=8.5cm\epsfysize=7.0cm
                     \epsfbox{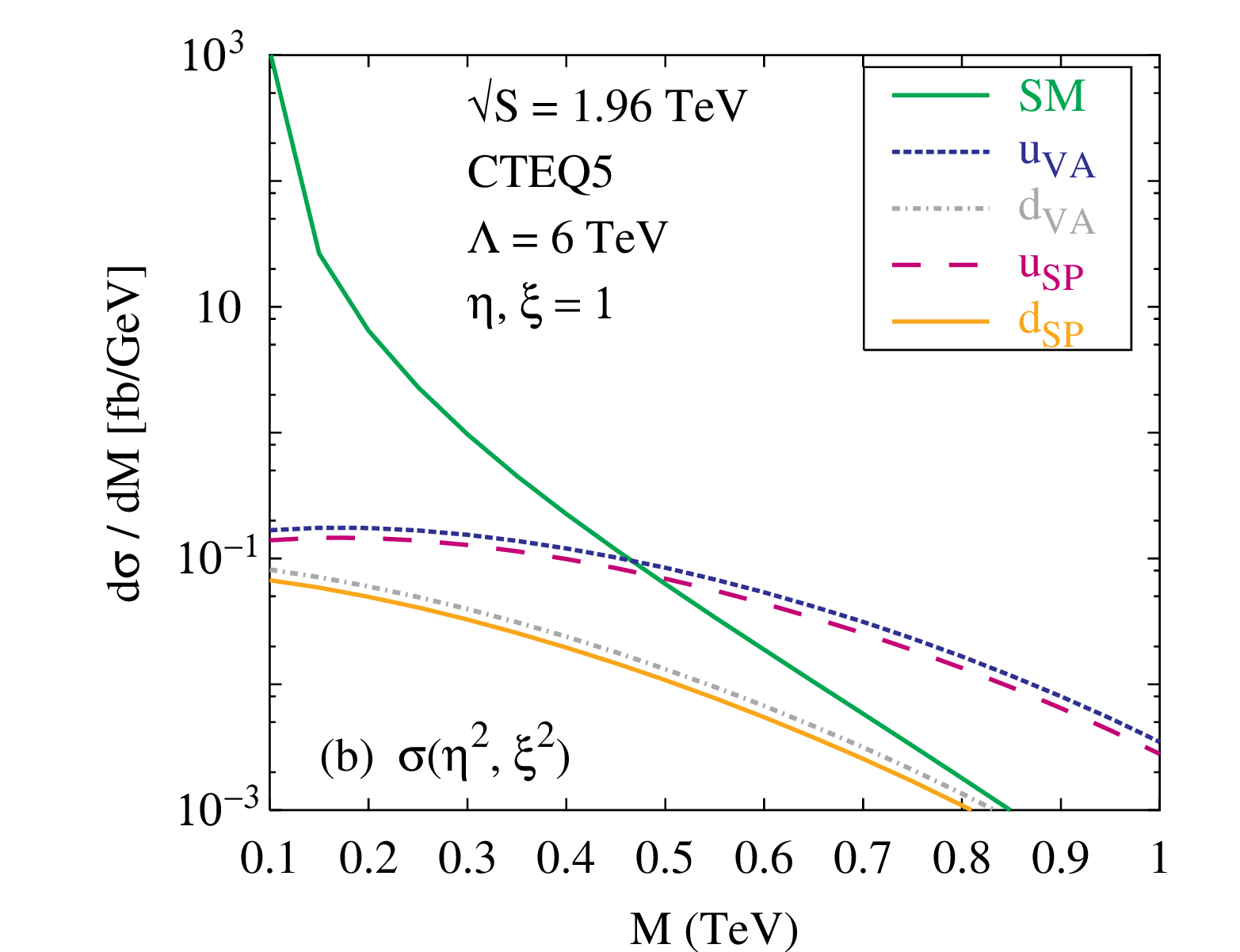}
}

\centerline{
\epsfxsize=8.5cm\epsfysize=7.0cm
                     \epsfbox{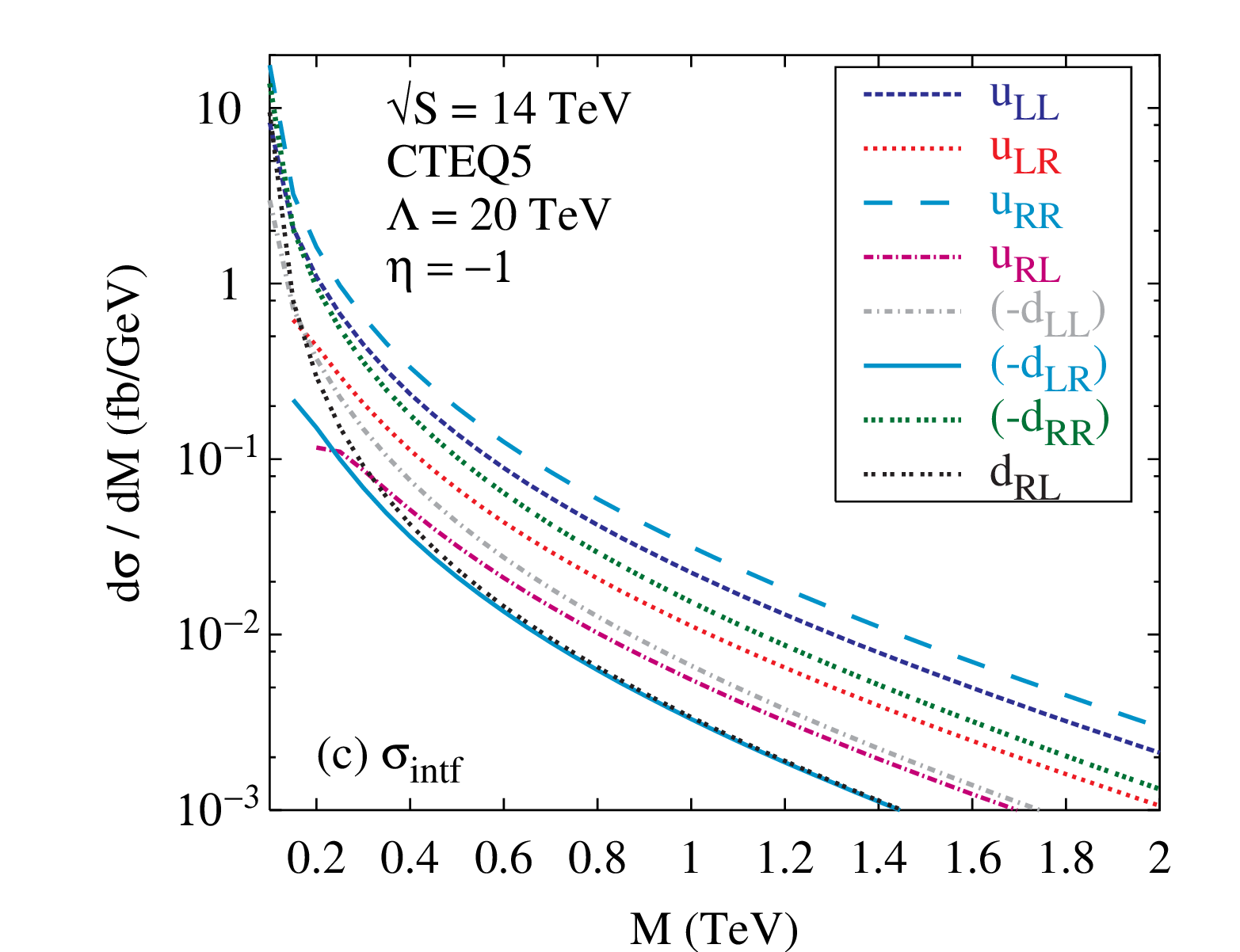}
\hspace*{-1em}
\epsfxsize=8.5cm\epsfysize=7.0cm
                     \epsfbox{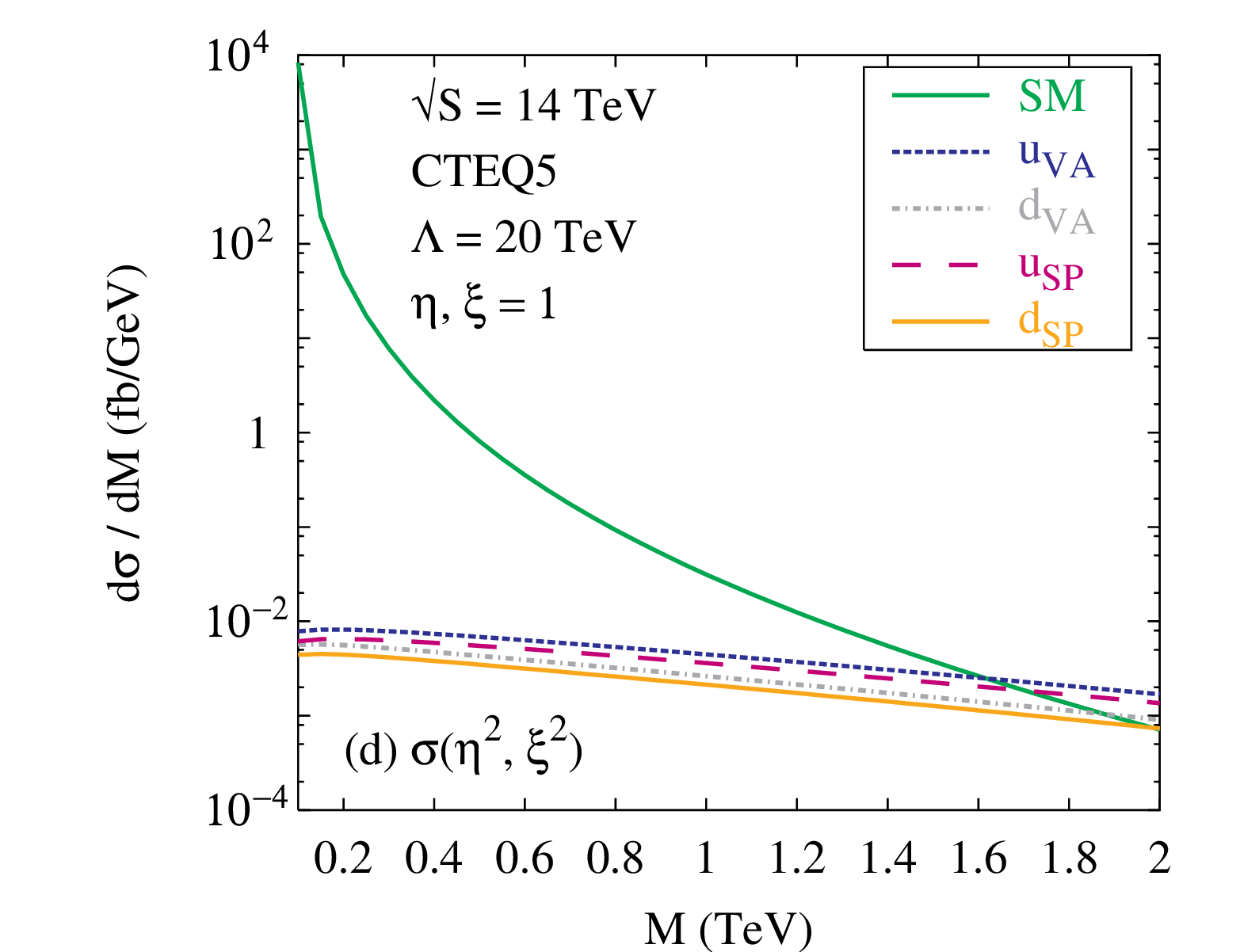}
}
\caption{\em The differential inclusive dilepton production
cross-sections (at NLO) for the contact interaction terms. 
In each case, only one coupling ($\eta, \xi$) is assumed to
be non-zero and of unit size. Also shown, for comparison, is the total
SM contribution.  The top and bottom panels refer to the Tevatron
($\sqrt{S} = 1.96$ TeV and $\Lambda = 6$ TeV) and the LHC ($\sqrt{S} =
14$ TeV and $\Lambda = 20$ TeV) respectively. The right and left 
panels  refer respectively to the pure contact interaction
term and the interference with the SM.}
\label{fig:csec_mdep}
\end{figure}

In Fig.\ref{fig:csec_mdep}, we present the invariant mass
distributions corresponding to the $VA$ case. Note that the
$\sigma_{\eta^2}$ piece (and, similarly, the $\sigma_{\xi^2}$ piece)
depends only on the identity of the quark $q$ taking part in the
contact interaction and is independent of the chirality structure of
the coupling. The interference term, on the other hand, does depend on
the chirality structure as Fig.\ref{fig:csec_mdep}($a$) amply
demonstrates.  As for $\sigma_{\rm SM}$, the rapid decrease in cross
section with $M$ is reflective of both the $s^{-1}$ fall
of the parton-level cross section as well as the rapid fall in parton
distribution functions at higher momentum fractions. That the
interference terms do not fall as fast is a consequence of the higher
dimensional nature of the contact interaction Lagrangian. This,
naturally, is even more evident for the $\sigma_{\eta^2}$
($\sigma_{\xi^2}$) piece. Consequently, at high $M$ values, 
the contact interaction contribution dominates over 
the SM piece. For the LHC, this dominance  occurs 
at a larger $M$ value as compared to the case of the
Tevatron precisely because we have chosen to work with a much 
larger value of $\Lambda$ for the former environment. And, expectedly, 
for identical couplings, 
the cross section due to a $ u \bar{u}$ initial state dominates that 
originating from a $ d \bar{d}$ initial state. In Fig.\ref{fig:csec_mdep},
we have chosen to limit ourselves only to these two initial states
as the cross sections corresponding to the heavier quarks would be 
suppressed even further (note though that the experimental bounds on 
$\Lambda$ is relaxed too for such cases). 

\begin{figure}[!h]
\centerline{
\epsfxsize=8.5cm\epsfysize=7.0cm
                     \epsfbox{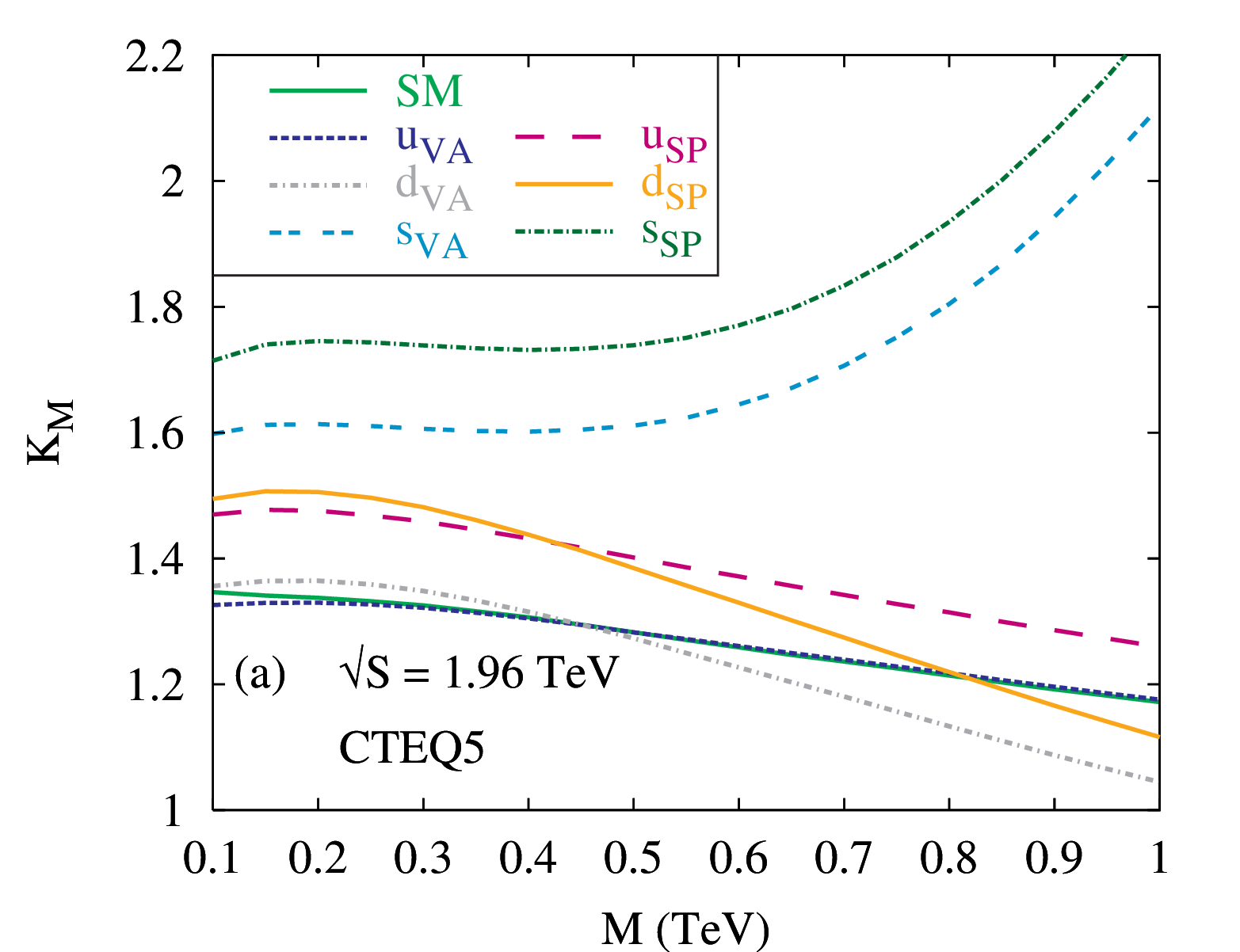}
\hspace*{-2em}
\epsfxsize=8.5cm\epsfysize=7.0cm
                     \epsfbox{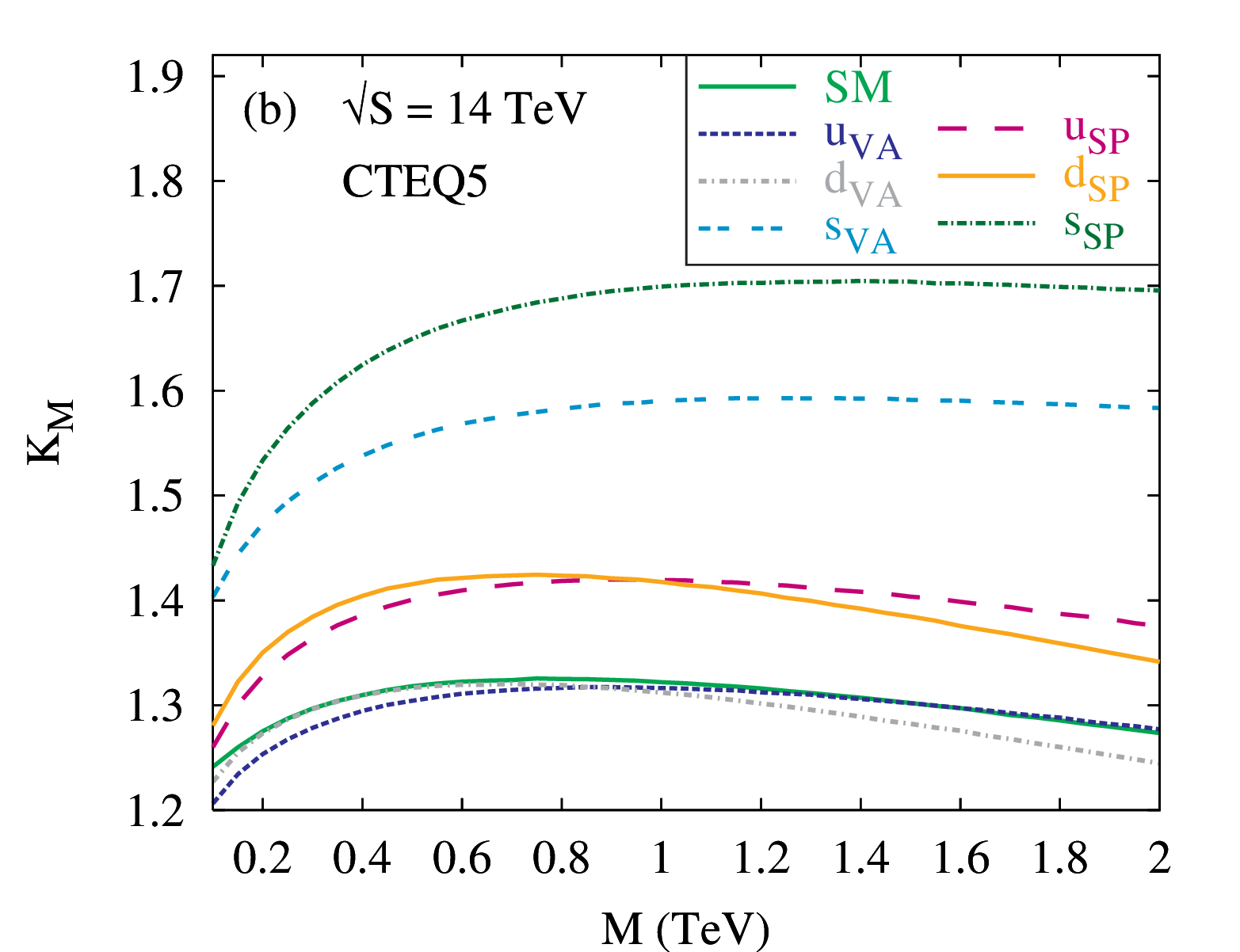}
}
\caption{\em The $K$-factors for the differential (in dilepton invariant mass)
cross-section for {\em (a)} the Tevatron Run II and {\em (b)} the LHC. 
For the contact interactions, the $K$-factors are independent of the 
chirality structure of the operators, but depend on whether they are of 
the $VA$ or the $SP$ type.
        }
\label{fig:kfac_mdep}
\end{figure}

We should clarify, at this stage, that, in calculating the NLO 
cross sections shown in Fig.\ref{fig:csec_mdep}, we have made a 
specific choice of the renormalization scale $\mu_R$ and the 
factorization scale $\mu_F$, namely,
\[ 
\mu_R = \mu_F = M \equiv \sqrt{Q^2}\ .
\]
Postponing, until later, a discussion of the dependence on the 
scale choice, we may define now a invariant mass-dependent $K$-factor, 
namely 
\beq
K^q_M = \Bigg[ {d \sigma_{LO}(M) \over dM} \Bigg]^{-1}
       \Bigg[ {d \sigma_{NLO}(M) \over dM} \Bigg] \,,
\eeq
where $q$ refers to the identity of the quark, 
and the LO (NLO) cross sections are computed by convoluting 
the corresponding parton-level cross sections with the LO (NLO) 
parton distribution functions. In Fig.\ref{fig:kfac_mdep}, 
we exhibit the variation of $K_M$ with $M$ for different 
choices of $q$.

As derived in the previous section, and as already 
evinced in Fig.\ref{fig:csec_mdep}, the fractional correction 
depends only on the spin structure of the vertex, and not 
on the chirality. Thus, for a given quark, the $K$-factor would depend 
on whether the interaction is of $SP$ or $VA$ type, but within each class,
the chirality structure (namely whether it is $LL$, $RR$, $LR$ or $RL$ 
type) is quite irrelevant. The last statement also implies that, for the 
$VA$-type interaction, the $K$-factor would be exactly the same as in 
the $SM$, as far as the particular quark initial state is concerned. 
Numerically, this feature is displayed in Fig.\ref{fig:kfac_mdep}. 
Of course, the $K$-factor {\em does} depend on the identity of $q$. As can be 
expected, $K_M^u(VA)$ and $K_M^d(VA)$ are relatively close to each other 
and, in turn, to $K_M(SM)$. In fact, to a large measure,  $K_M(SM)$ is but 
the weighted average of the other two, with the relative strengths being 
determined by the quark fluxes. That these $K$-factors fall monotonically 
with $M$ for the case of the Tevatron and not so for the 
LHC is understandable in the light of the fact that, the former 
is a $\overline p p$ machine, while the latter is a $p p $ one. As for 
$K_M^s(VA)$, the steep rise at large  $M$ values is but 
a reflection of the dominance of the Compton-like subprocess 
($s g \to \ell^+ \ell^- s$ and $\bar s g \to \ell^+ \ell^- \bar s$) 
owing to the larger flux of gluons as compared to $s / \bar s$, especially 
for large momentum fractions. 

The results for the $SP$-type interactions are qualitatively similar, 
though quantitatively the $K$-factors are significantly larger than 
those for the  $VA$-type interaction (or the SM). 
The numerical differences are but consequences of the 
the structures of  the respective matrix elements. On closer inspection, 
$K^q_M(SP)$, for a given  $M$,  turns out to be the 
same as that for resonance production of a scalar/pseudoscalar 
of mass $M$~\cite{1st_paper}.

\subsection{The rapidity distributions}

We now turn to the distribution in a different kinematical variable, 
namely $Y$, {\em the rapidity of the lepton 
pair}\footnote{This should be distinguished from the rapidity of 
an individual lepton.}. However, rather than look at $d \sigma / d Y$
itself, we shall rather consider  on 
$d^2 \sigma / d M \, d Y$, for this 
allows us to accentuate the effect of the contact interactions by 
concentrating on a suitable $M$ range. 

\begin{figure}[!h]
\centerline{
\epsfxsize=8.5cm\epsfysize=7.0cm
                     \epsfbox{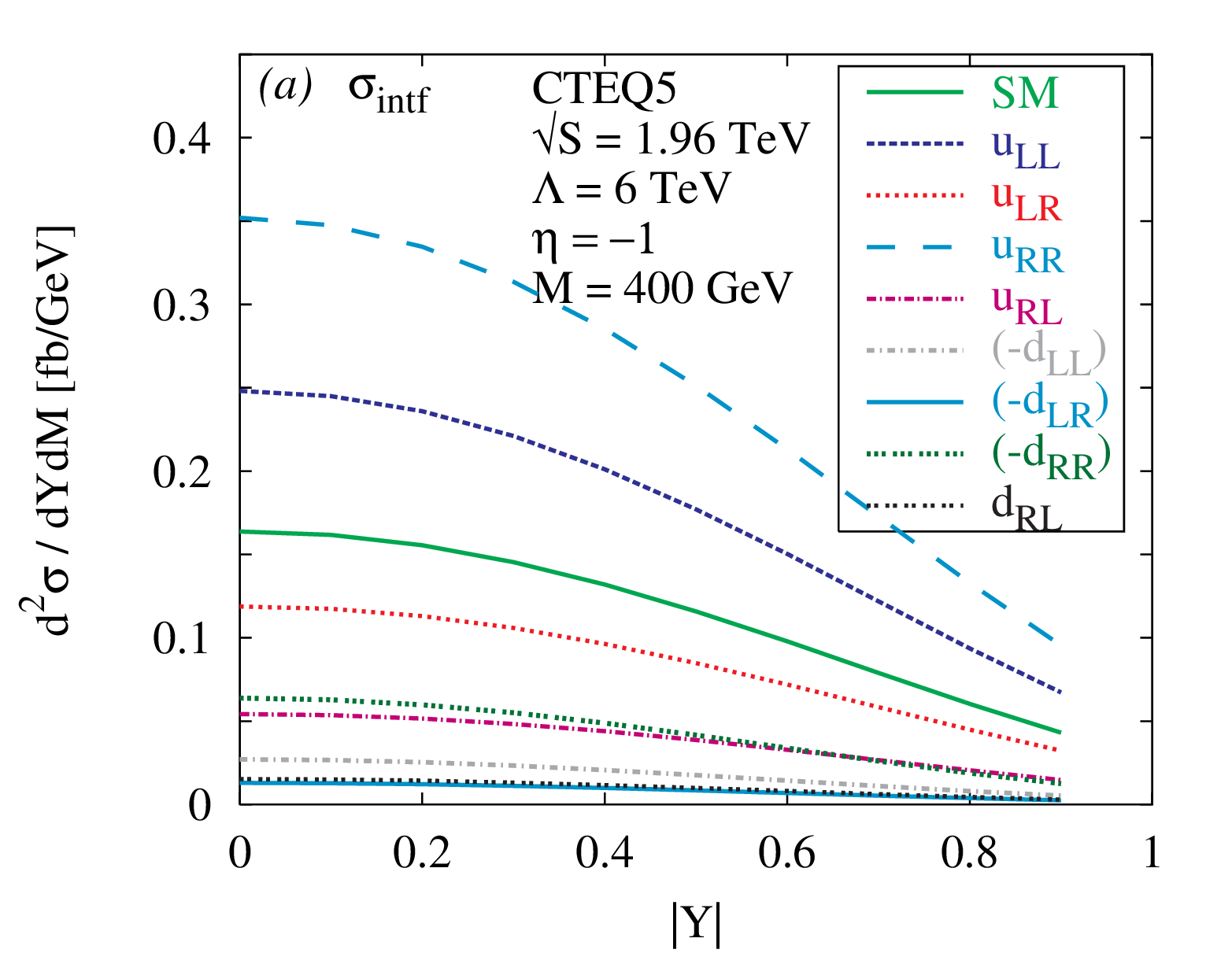}
\hspace*{-2em}
\epsfxsize=8.5cm\epsfysize=7.0cm
                     \epsfbox{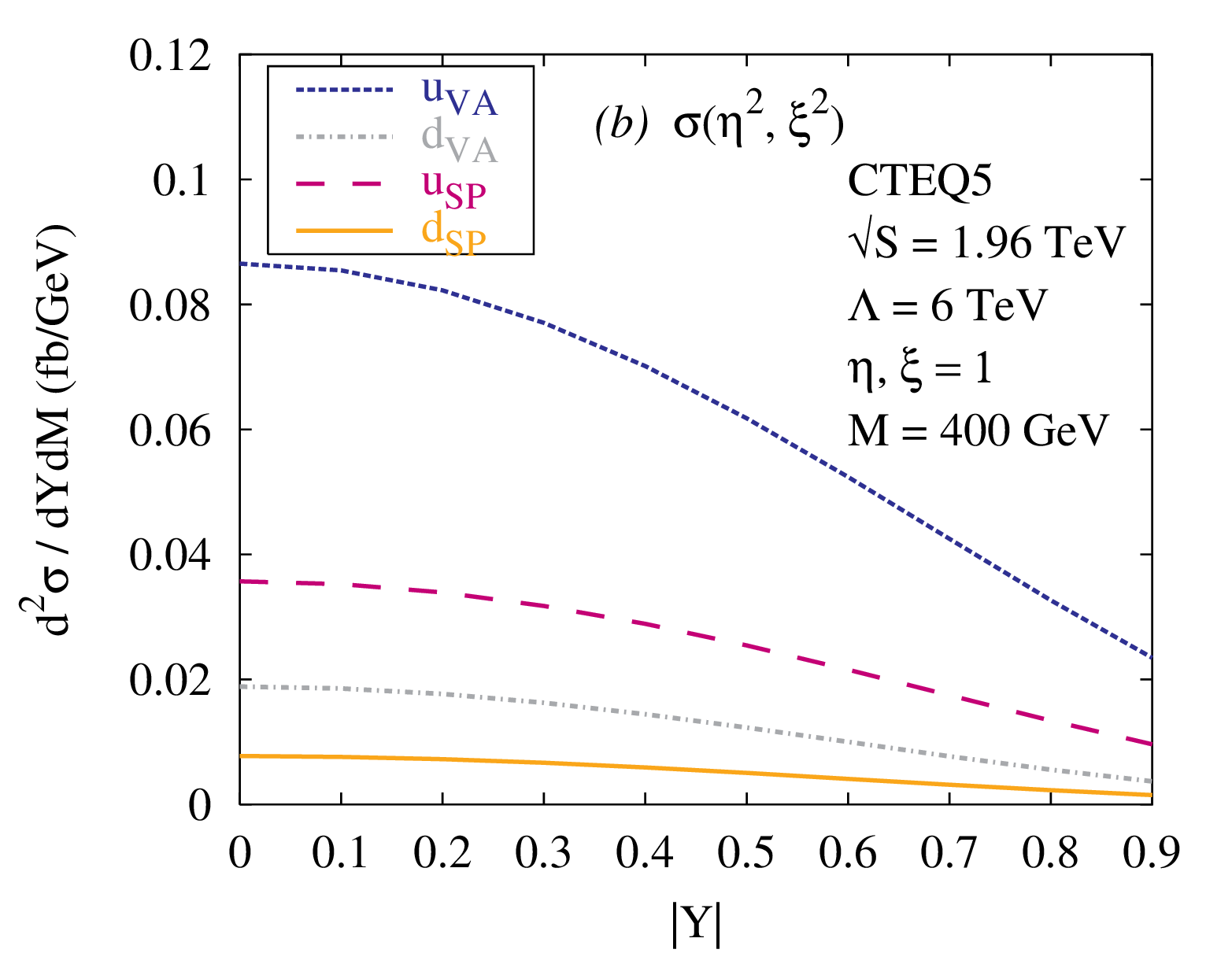}
}

\centerline{
\epsfxsize=8.5cm\epsfysize=7.0cm
                     \epsfbox{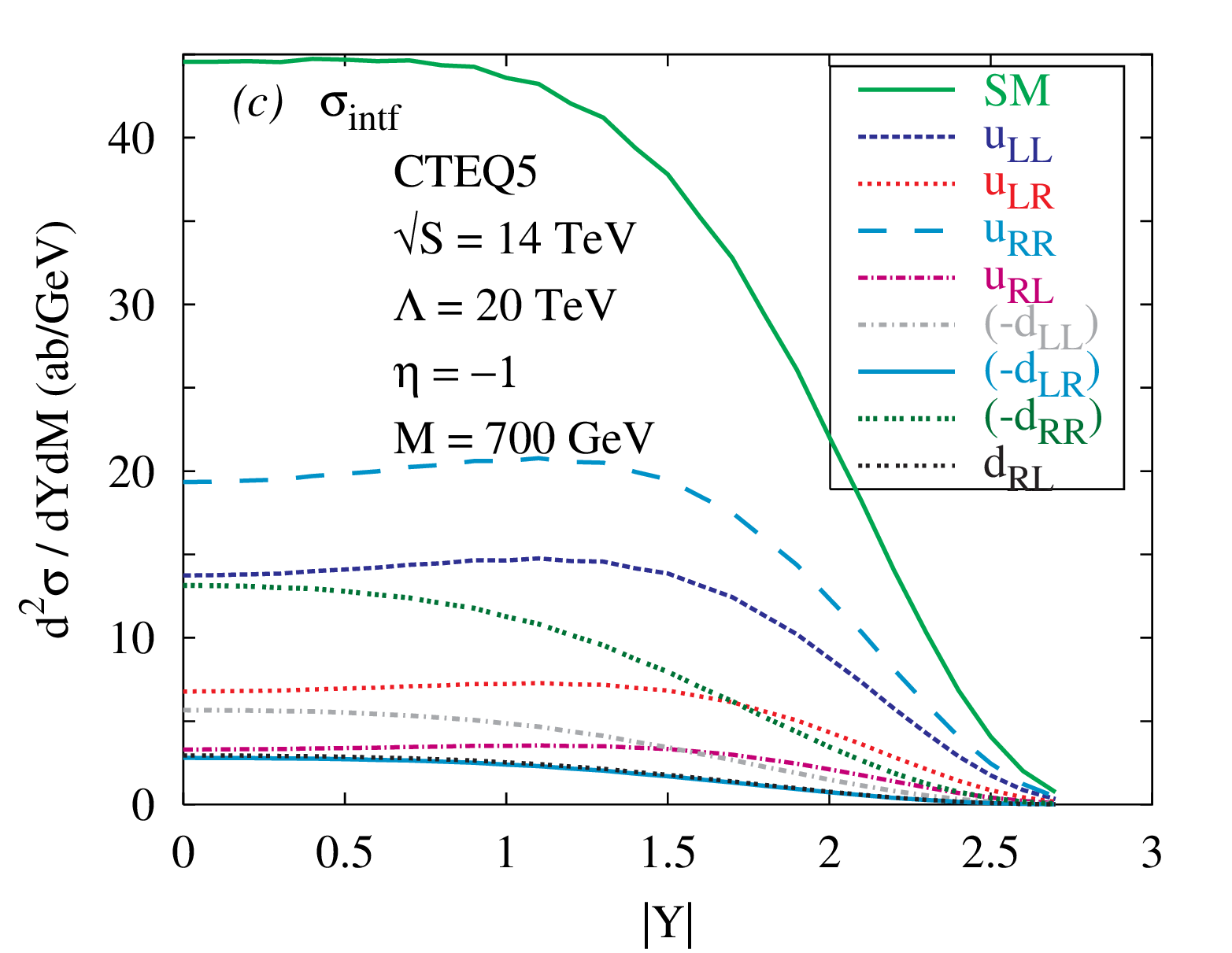}
\hspace*{-2em}
\epsfxsize=8.5cm\epsfysize=7.0cm
                     \epsfbox{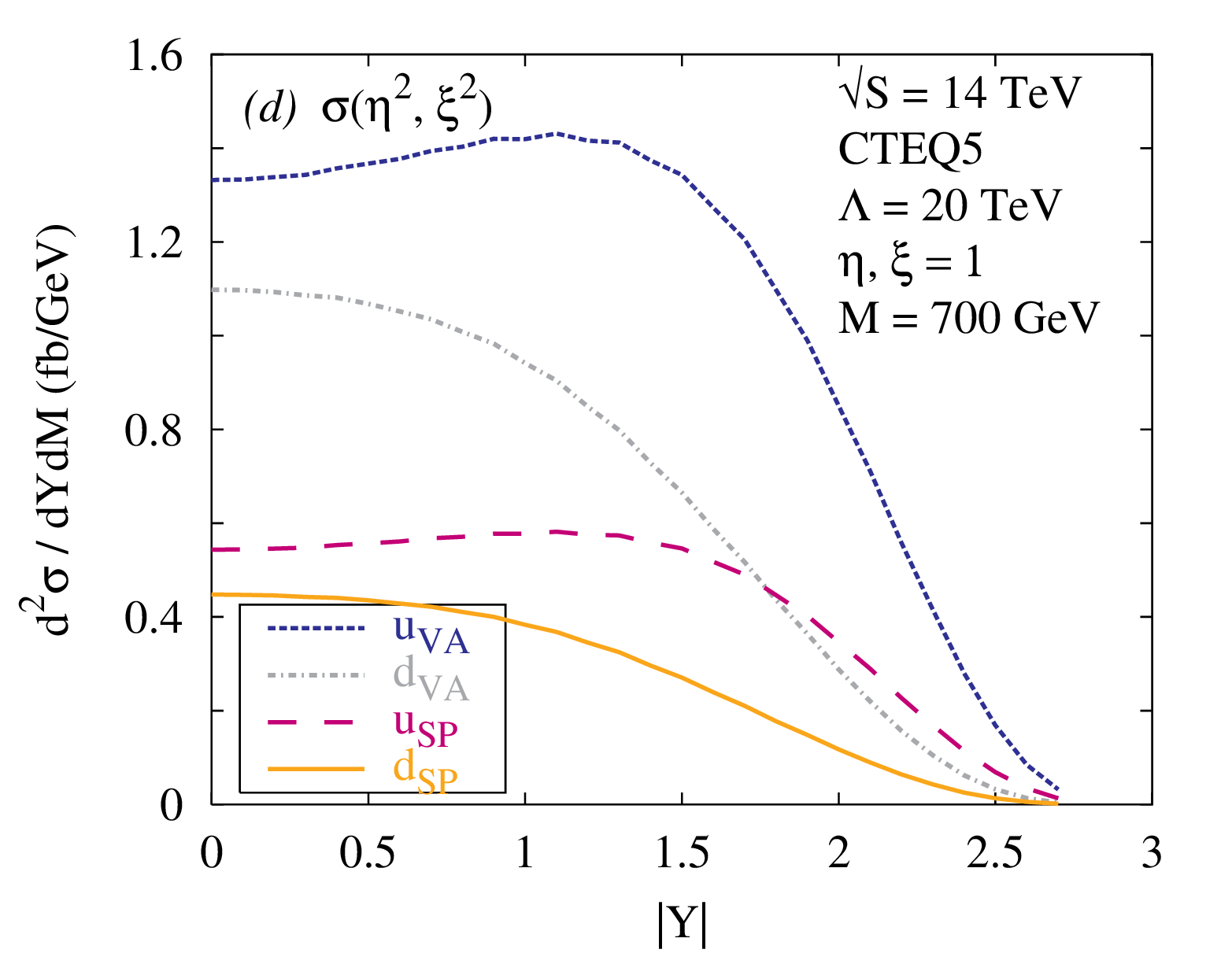}
}
\caption{\em As in Fig.\protect\ref{fig:csec_mdep}, but for the 
double differential (in rapidity $Y$ and mass $M$) distribution instead. 
        }
\label{fig:csec_rapdep}
\end{figure}

At the LO, the variable $Y$ is just a measure of the boost of the 
partonic center of mass with respect to the laboratory frame. 
At the NLO, one has to consider the effect of the initial state 
radiation as well. In either case, it is easy to see that 
$d^2 \sigma / d M \, d Y$ (and, hence,  $d \sigma / d Y$) 
is an even function of $Y$. In Fig.\ref{fig:csec_rapdep}, we exhibit 
the dependence of $d^2 \sigma / d M \, d Y$ on $Y$, for a 
fixed value of $M$. As expected, for a large enough value 
of the latter, the effect of the contact interaction is 
clearly discernible and especially for the central rapidity 
region. Note that the contact interaction cross sections are 
significantly flatter in $Y$ than the SM contribution. Once again, this 
is a reflection of the structure of the new physics matrix element 
as compared to that due to $\gamma / Z$ exchange. 

\begin{figure}[!h]
\centerline{
\epsfxsize=8.5cm\epsfysize=7.0cm
                     \epsfbox{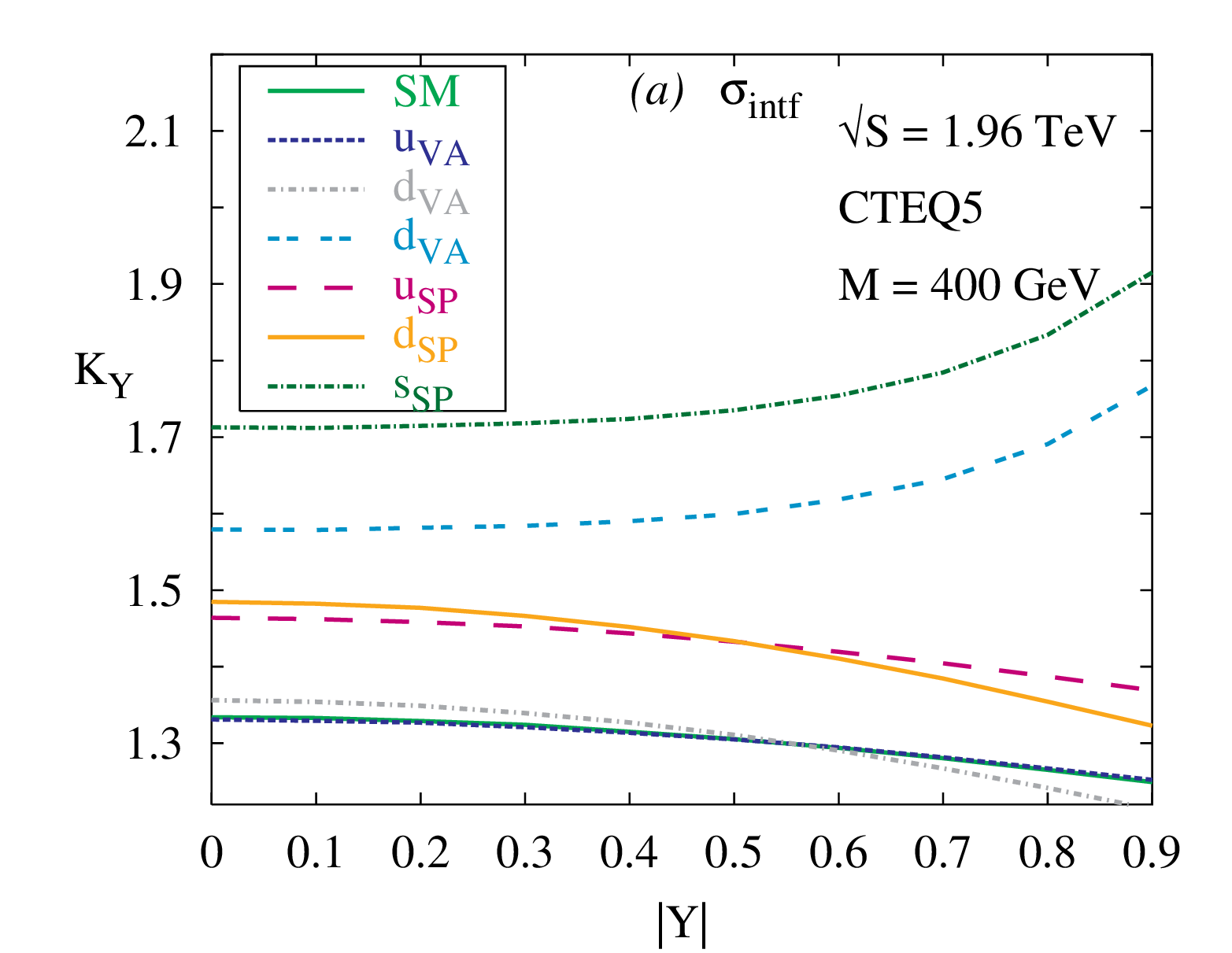}
\hspace*{-1em}
\epsfxsize=8.5cm\epsfysize=7.0cm
                     \epsfbox{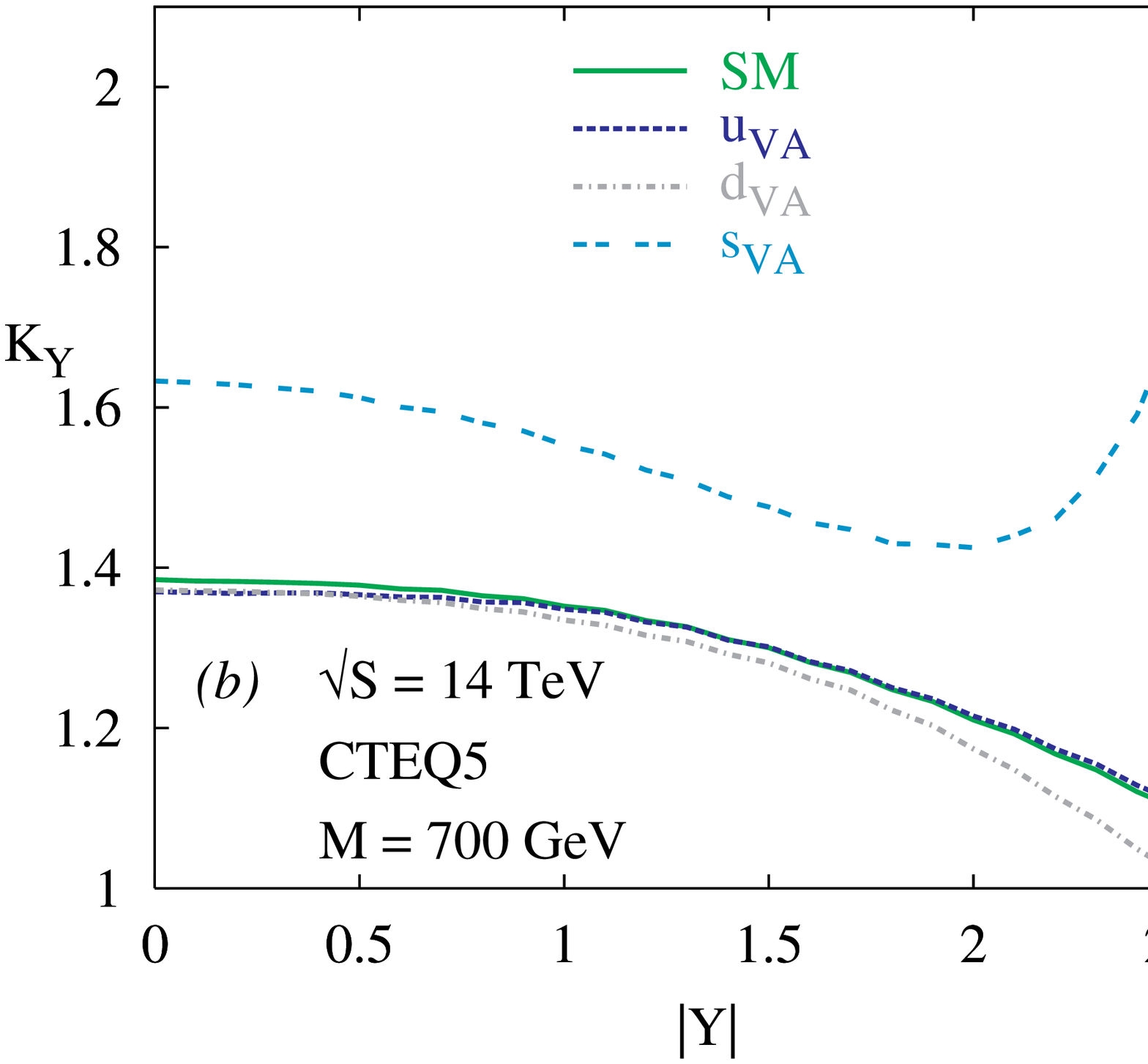}
}
\caption{\em  As in Fig.\protect\ref{fig:kfac_mdep}, but for the 
double differential (in rapidity $Y$ and mass $M$) distribution instead. 
        }
\label{fig:kfac_y}
\end{figure}

Analogous to $K_M$ defined in the previous subsection, one may now 
define a $Y$-dependent $K$-factor of the form 
defined as
\beq
K_Y \equiv \Bigg[ {d \sigma_{LO}(M,Y) \over {dM~dY}} \Bigg]^{-1}
       \Bigg[ {d \sigma_{NLO}(M,Y) \over {dM~dY}} \Bigg] \ ,
\eeq
and we plot this quantity as a function of $Y$ in Fig.\ref{fig:kfac_y}. 
The results are quite reminiscent of those for $K_M$ (as displayed 
in  Fig.\ref{fig:kfac_mdep}). It is noteworthy that, for the LHC, 
$K_Y^s$ shows a large upward swing at large $Y$, whereas $K_M^s$ had 
seemed better behaved. The reason is not difficult to fathom. For the 
range of $M$ spanned in Fig.\ref{fig:kfac_mdep}($b$), the 
cross section integral typically samples relatively moderate values 
of the Bjorken-$x$ as compared to the higher-$M_{\ell\ell}$ regime for 
the Tevatron case (Fig.\ref{fig:kfac_mdep}($a$)). On the other hand, 
a phase space point such as $(M = 700 \gev, |Y| = 2.5)$ 
necessarily pushes one to larger momentum fractions for the partons and 
thus, once again, it is the ratio of the strange-quark flux to that of the 
gluon that causes the upward turn in Fig.\ref{fig:kfac_y}($b$).

\subsection{The choice of scale}

Until now, we have chosen each of 
the factorization scale $\mu_F$ (relevant 
to both the LO as well as NLO calculations) and the renormalization scale
$\mu_R$ (relevant only for the NLO case) to be the same as the 
dilepton invariant mass $M$. As is well known, this choice 
is arbitrary and there is no theoretical guideline for making such 
a choice. Maintaining, for reasons of simplicity, $\mu_R = \mu_F$, we 
now examine the dependence of our calculations on this choice. To 
quantify the scale dependence of
our result, we define ratios $R_M$ and $R_Y$
\beq
\begin{array}{rcl}
 R^I_M (\mu_F) & \equiv & \dis \Bigg[ {d
\sigma_I(M,\mu_F = M) \over dM} \Bigg]^{-1} \; 
       \Bigg[ {d \sigma_I(M,\mu_F) \over dM} \Bigg] \,,
\\[3ex]
R^I_Y (\mu_F) 
& \equiv & \dis \Bigg[ {d \sigma_I(M,Y,\mu_F = M) \over {dM~dY}} \Bigg]^{-1}
     \;  \Bigg[ {d \sigma_I(M,Y,\mu_F) \over {dM~dY}} \Bigg],
\end{array}
\eeq
where $I =$~LO,~NLO. A value of $ R^I_{M, Y} (\mu_F)$ close to 
unity would then signify a low sensitivity to the choice of scale 
and hence a more robust result. 

\begin{figure}[!h]
\centerline{
\epsfxsize=8.8cm\epsfysize=9.0cm
                     \epsfbox{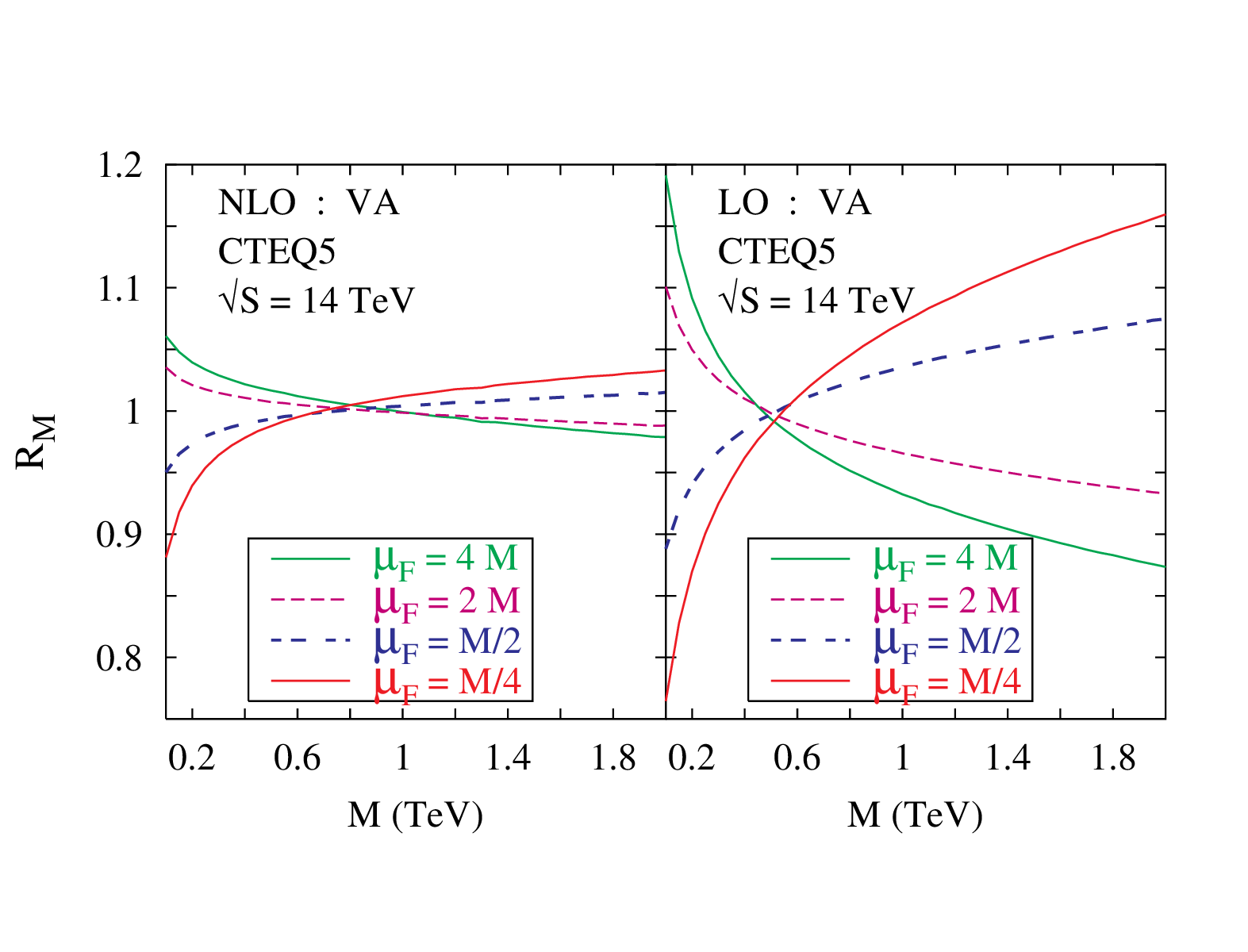}
\hspace*{-1em}
\epsfxsize=8.8cm\epsfysize=9.0cm
                     \epsfbox{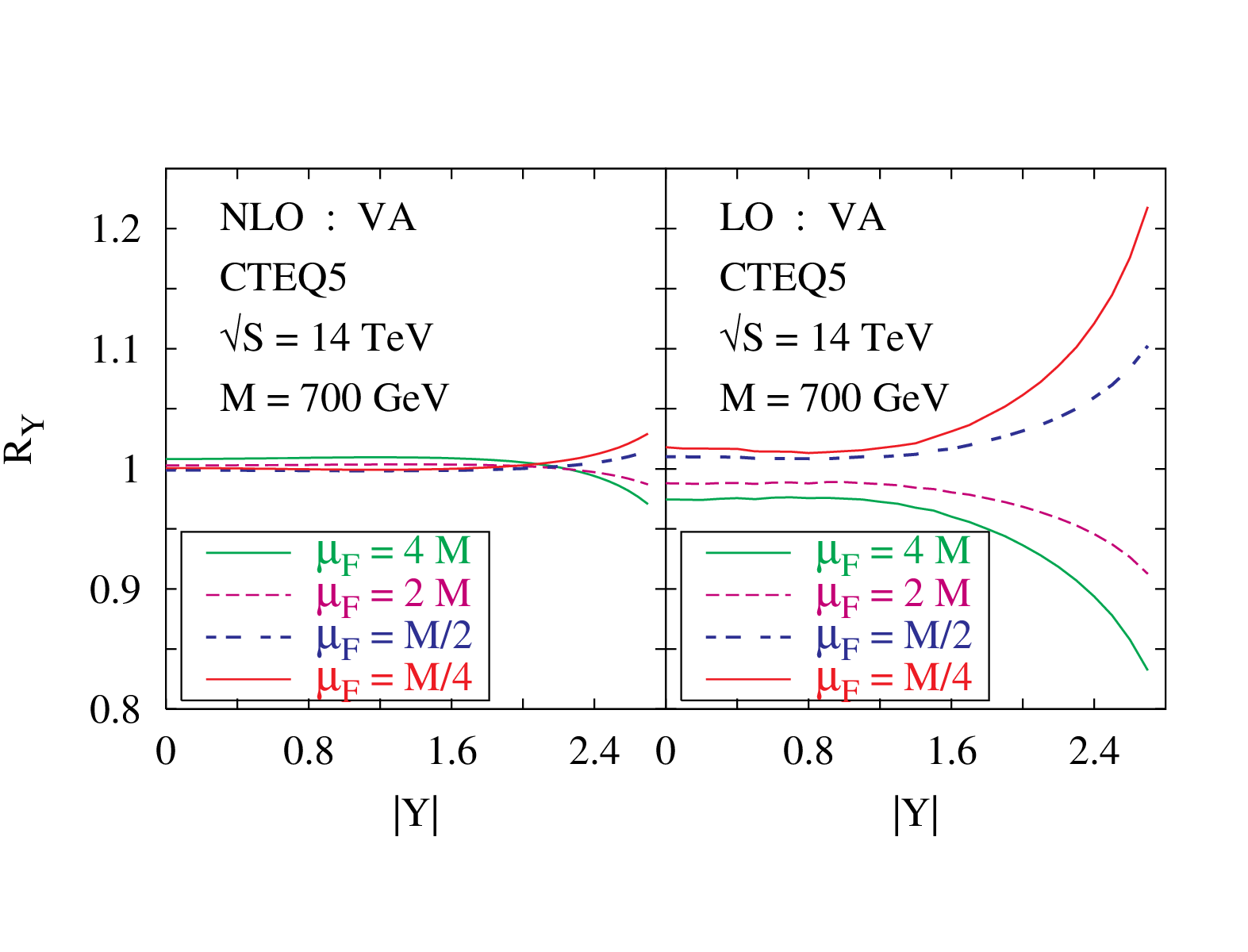}
}
\caption{\em The $K$-factors for the differential (in dilepton invariant mass)
cross-section for {\em (a)} the Tevatron Run II and {\em (b)} the LHC. 
For the contact interactions, the $K$-factors are independent of the 
chirality structure of the operators, but depend on whether they are of 
the $VA$ or the $SP$ type.
        }
\label{fig:scal_VA}
\end{figure}

In Fig.\ref{fig:scal_VA},  we display the above ratios 
for the case of the LHC and the $VA$ interactions. 
Note that the variation of the cross
section with the factorization scale is relatively small.
Furthermore, the variation reduces significantly 
as one goes from the $LO$ to the 
$NLO$ case. This immediately points to the increased 
robustness of the prediction on inclusion of the 
corrections, and lends hope that the remaining scale
ambiguity can, presumably, be reduced by adding still higher order
corrections. Note that, at the leading order, these ratios 
are independent of the dynamics and reflect only the effect 
of the choice of the factorization scale on the parton densities. 
In other words,
\[
R^{LO}_M (SP) = R^{LO}_M (VA), \qquad
R^{LO}_Y (SP) = R^{LO}_Y (VA) \ .
\]
At the next-to-leading order, the dynamics does play a role. However, the 
differences between the $R$-ratios for the $SP$ and $VA$ cases are too small 
to be noticeable on the scale of  Fig.\ref{fig:scal_VA}. The results 
are similar for the case of the Tevatron as well.

\section{Conclusions}

To summarize, we have performed a systematic calculation of the 
next-to-leading order QCD corrections for the Drell-Yan process
in theories with contact 
interactions. Contrary to naive expectations, we demonstrate explicitly 
that the QCD corrections are meaningful and reliable in the sense that 
no undetermined parameters need be introduced.

We have analyzed both the invariant mass distribution and the 
rapidity distributions for the dilepton pair at either of the 
Tevatron and the LHC. The enhancements over the LO expectations are 
found to be quite significant. The corresponding $K$-factors are presented 
in a form suitable for use in experimental analyzes. 

For the $VA$-type interactions, the analytical structure of the corrections 
are similar to those for the SM. However,  a significant dependence 
on the flavour structure is found and needs to be carefully 
accounted for in obtaining any experimental bounds. For the $SP$-type 
interaction, not only are the analytical results  quite different, but the 
consequent $K$-factors are typically larger than those within the SM. 

Finally, we have investigated the sensitivity of our results to 
both the factorization and renormalization scales. As expected, we 
find such dependences to be greatly reduced for the case of the NLO 
results as compared to that for the LO case. This indicates the robustness 
of the calculations. 

\section*{Acknowledgements} 
DC thanks the
Department of Science and Technology, India for financial assistance
under the Swarnajayanti Fellowship grant. The work of SM is partly supported
by an US DOE grant No. DE-FG02-05ER41399 and NATO grant No. PST-CLG. 980342.

\end{document}